\documentclass[final,5p,times,twocolumn]{elsarticle}
\usepackage{amssymb}
\usepackage{amsmath,amsfonts}
\usepackage{bm}
\usepackage{nomencl}
\makenomenclature
\usepackage{algorithmic}
\usepackage{algorithm}
\usepackage{array}
\usepackage{subfig}
\usepackage{xcolor}
\newcolumntype{M}[1]{>{\centering\arraybackslash}m{#1}}
\usepackage{array}
\usepackage{multicol}
\usepackage{multirow}
\usepackage{url}
\usepackage{graphicx}
\usepackage{bm}
\usepackage{comment}

\begin{document}

\begin{frontmatter}

\title{Deep Reinforcement Learning for Wind and Energy Storage Coordination in Wholesale Energy and Ancillary Service Markets}

\author[label1]{Jinhao Li}
\ead{stephlee175@gmail.com}
\author[label2,label3]{Changlong Wang}
\ead{chang.wang@monash.edu}
\author[label1,label3]{Hao Wang\corref{cor1}}
\ead{hao.wang2@monash.edu}
\cortext[cor1]{Corresponding author: Hao Wang.}
\affiliation[label1]{
            organization={Department of Data Science and AI, Faculty of Information Technology, Monash University},
            % addressline={},
            city={Melbourne},
            % postcode={3800}, 
            state={Victoria},
            country={Australia}
}
\affiliation[label2]{
            organization={Department of Civil Engineering, Monash University},
            % addressline={},
            city={Melbourne},
            % postcode={3800}, 
            state={Victoria},
            country={Australia}
}
\affiliation[label3]{
            organization={Monash Energy Institute, Monash University},
            % addressline={},
            city={Melbourne},
            % postcode={3800}, 
            state={Victoria},
            country={Australia}
}

\begin{abstract}
%% Text of abstract
Wind energy has been increasingly adopted to mitigate climate change. However, the variability of wind energy causes wind curtailment, resulting in considerable economic losses for wind farm owners. Wind curtailment can be reduced using battery energy storage systems (BESS) as onsite backup sources. Yet, this auxiliary role may significantly weaken the economic potential of BESS in energy trading. Ideal BESS scheduling should balance onsite wind curtailment reduction and market bidding, but practical implementation is challenging due to coordination complexity and the stochastic nature of energy prices and wind generation. We investigate the joint-market bidding strategy of a co-located wind-battery system in the spot and Regulation Frequency Control Ancillary Service markets. We propose a novel deep reinforcement learning-based approach that decouples the system's market participation into two related Markov decision processes for each facility, enabling the BESS to absorb onsite wind curtailment while performing joint-market bidding to maximize overall operational revenues. Using realistic wind farm data, we validated the coordinated bidding strategy, with outcomes surpassing the optimization-based benchmark in terms of higher revenue by approximately 25\% and more wind curtailment reduction by 2.3 times. Our results show that joint-market bidding can significantly improve the financial performance of wind-battery systems compared to participating in each market separately. Simulations also show that using curtailed wind generation as a power source for charging the BESS can lead to additional financial gains. The successful implementation of our algorithm would encourage co-location of generation and storage assets to unlock wider system benefits.
\end{abstract}

\begin{comment}
%%Graphical abstract
\begin{graphicalabstract}
\includegraphics[width=0.8\linewidth]{figs/graph_abstract.001.png}
\end{graphicalabstract}

%%Research highlights

\begin{highlights}
    \item Synergy between reduced wind curtailment and market bidding for wind-battery systems
    \item Deep reinforcement learning-based bidding strategy in spot and regulation markets
    \item Our strategy significantly outperforms optimization-based benchmark 
    \item Storing curtailed wind energy in onsite batteries could boost financial returns    
\end{highlights} 
\end{comment}

\end{frontmatter}

\nomenclature[01]{$v_t^\text{W}$}{Proportion of wind farm bid power in the spot market}
\nomenclature[02]{$v_t^\text{Dch}$}{Binary variable indicating the BESS discharge operation}
\nomenclature[03]{$v_t^\text{Ch}$}{Binary variable indicating the BESS charge operation}
\nomenclature[04]{$p_t^\text{W}$}{Forecasted wind generation availability}
\nomenclature[05]{$p_t^\text{W,Act}$}{Actual wind generation}
\nomenclature[06]{$p_t^\text{W,WC}$}{Curtailed wind generation}
\nomenclature[07]{$p_t^\text{BESS,S}$}{Battery bid power in the spot market}
\nomenclature[08]{$p_t^\text{BESS,Reg}$}{Battery bid power in Regulation FCAS market}
\nomenclature[09]{$p_t^\text{BESS,WC}$}{Power that BESS plan to draw from wind curtailment}
\nomenclature[10]{$\hat{p}_t^\text{BESS,WC}$}{Actual power drawn by BESS from onsite wind farm}
\nomenclature[11]{$\bm{s}_t$}{A series of AGC signals in one dispatch interval}
\nomenclature[12]{$\rho_t^\text{S}$}{Market clearing price in the spot market}
\nomenclature[13]{$\rho_t^\text{RR}$}{Market clearing price (i.e., ancillary service price) in Regulation Raise FCAS market}
\nomenclature[14]{$\rho_t^\text{RL}$}{Market clearing price (i.e., ancillary service price) in Regulation Lower FCAS market}
\nomenclature[15]{$e_t$}{Battery current capacity}
\nomenclature[16]{$L$}{Length of AGC signals in one dispatch interval}
\nomenclature[17]{$\Delta s$}{Duration of each AGC signal}
\nomenclature[18]{$\Delta t$}{Duration of one NEM dispatch interval.}
\nomenclature[19]{$T$}{The overall time frame}
\nomenclature[20]{$\lambda$}{Penalty coefficient for deviations between actual wind generation and prescribed dispatch target}
\nomenclature[21]{$\eta^\text{Dch}$}{Battery discharging efficiency}
\nomenclature[22]{$\eta^\text{Ch}$}{Battery charging efficiency}
\nomenclature[23]{$c$}{Battery degradation cost coefficient}
\nomenclature[24]{$P_\text{max}^\text{W}$}{Wind installed capacity}
\nomenclature[25]{$P_\text{max}^\text{BESS}$}{Battery rated power}
\nomenclature[26]{$E_\text{min}$}{Battery lower energy storage limit}
\nomenclature[27]{$E_\text{max}$}{Battery upper energy storage limit}
\printnomenclature

%% main text
\section{Introduction} \label{sec:intro}
In the past decade, wind energy has played a major role in decarbonizing power systems and addressing climate change through the transition to net-zero emissions~\cite{IPCC2022}. In Australia, wind energy accounts for $9.9$\% of total electricity production~\cite{CEC2022}, making it the leading source of renewable energy at the utility scale. Currently, there are $9.7$ GW of wind farms operating in the Australian National Electricity Market (NEM), with an additional $77$ GW planned for construction over the next decade~\cite{aemo_future_gen}. However, the variable nature of wind and lack of accurate wind forecasts make it difficult to accommodate and dispatch wind power in real-time, leading to inevitable curtailments by the market operator to ensure system security and reliability~\cite{burke2011} at the cost of wind producers. As wind power adoption has increased, so has wind curtailment, resulting in significant losses for wind power producers. Similarly, the adoption of battery energy storage systems (BESS) has also seen growth, with approximately $650$ MW of batteries registered in the NEM and an additional $34.3$ GW planned for the next decade~\cite{aemo_future_gen}.

The co-location of renewable energy and battery energy storage systems is becoming increasingly common, as coordinated investment in these technologies can reduce curtailment, diversify revenue streams, mitigate market risks, and delay the need for network expansion. This is supported by the findings of the Australian Energy Market Operator (AEMO), which recommends co-location of renewable energy and BESS in dedicated renewable energy zones in energy system planning~\cite{AEMOSysPlan2022}. As the adoption of wind and BESS technologies continues to grow rapidly, it is crucial to understand how to effectively coordinate these technologies for the benefit of the power grid and their financial performance, as part of the transition towards a sustainable energy future.

The cost-effectiveness of storing curtailed wind energy depends on the upfront costs of storage and the underlying coordination strategy between generation and storage assets. When co-located, the BESS can serve as a storage medium to reduce wind curtailment by shifting surplus wind generation to periods of low production. For instance, \cite{sun2017} proposed a scenario-based stochastic optimization method to reduce wind curtailment while \cite{dui2018} presented a two-stage optimization approach and \cite{sun2017} developed a linear programming-based framework to optimize the power and storage capacity of the BESS for wind curtailment management. Previous research has shown that optimal sizing and operation of BESS under this configuration is typically determined through stochastic or robust optimization methods, which require knowledge of wind power uncertainty distributions. However, this auxiliary role may limit the BESS' ability to profit from participation in electricity markets, particularly through energy arbitrage in the wholesale spot market (i.e., buy low and sell high) and provision of frequency control ancillary services (FCAS) in FCAS markets, which are critical sources of profitability for utility-scale BESS in the current electricity system. 

Optimization-based methods have also been applied in the study of coordination between wind energy and battery energy storage systems (BESS) in electricity market bidding strategies. For example, \cite{zhang2018} proposed a market-oriented optimal dispatching strategy for a wind farm equipped with a hybrid BESS. \cite{akbari2019,xie2021} developed stochastic and robust optimization models for wind-battery systems participating in the electricity market, respectively. These studies typically consider wind farms and BESS as separate entities bidding in an aggregated manner, without addressing wind curtailment management in the BESS bidding process. The performance of these strategies is also heavily influenced by energy price forecasts, which can be difficult to accurately predict due to the volatile nature of the electricity market and the complexity of price drivers~\cite{weron2014}.

There has been a lack of research on real-time bidding strategies for co-located wind and BESS using methods other than optimization-based approaches. Our study addresses this gap by developing a deep reinforcement learning (DRL)-based bidding strategy for co-located wind-BESS systems. This strategy allows the wind-BESS system to simultaneously reduce wind curtailment and maximize revenue through joint bidding in spot and regulation FCAS markets. Unlike model-based approaches, the DRL approach is model-free and can learn uncertainties associated with wind generation and energy prices from historical observations without requiring prior knowledge or energy price forecasts. Additionally, the DRL allows the wind-BESS system to dynamically balance the trade-off between market participation and wind curtailment mitigation in real-time trading. For simplicity, we refer to our method, the aggregated bidding of the co-located wind-BESS system in the joint market via deep reinforcement learning, as ``AggJointDRL''. The main contributions of our study are summarized below.
\begin{itemize}
    \item \emph{Synergizing Wind Curtailment Management and Market Bidding}: We examine the synergies between wind curtailment management and real-time market bidding in a co-located wind-battery system, highlighting the importance of flexible and dynamic coordination between generation and storage for profitability.
    \item \emph{DRL-based Joint-Market Bidding}: We study the optimal joint-market bidding strategy of the wind-battery system in both the spot and regulation FCAS markets, utilizing a model-free DRL algorithm -- the twin delayed deep deterministic policy gradient (TD3) -- to optimize market participation of the wind-battery system through two related Markov decision processes (MDP).
    \item \emph{Numerical Simulations and Implications}: We validated our AggJointDRL using realistic wind farm data from the Australian National Electricity Market. Our results show that our method excels in two areas: 1) generating significant revenue increases of approximately $25\%$ compared to the optimization-based benchmark; and 2) fully leveraging the flexibility in both the spot and FCAS markets to improve the overall joint bidding outcome, which is significantly more profitable than individual market participation. Effective coordination between wind and battery can also boost financial returns by charging the BESS with curtailed wind energy.
\end{itemize}

The rest of the paper is organized as follows. Section \ref{sec:literature} reviews the related work; Section \ref{sec:model} formulates the joint-market bidding problem of the wind-battery system; Section \ref{sec:method} decouples the aggregated bidding of the co-located wind-battery system into two MDPs and introduce the DRL to concurrently maximize the overall revenue and reduce wind curtailment; Section \ref{sec:exp} presents and discusses simulation results; and Section \ref{sec:conclusions} concludes.

\section{Related Works} \label{sec:literature}
Several studies have focused on optimizing the use of battery energy storage systems (BESS) to minimize wind curtailment. For example, \cite{sun2017} proposed a scenario-based stochastic optimization approach to reduce wind curtailment and minimize operating costs in the power system, while \cite{dui2018} presented a two-stage optimization approach and \cite{alanazi2017} developed a linear programming-based framework to optimize the power and storage capacity of the BESS for wind curtailment management. Additionally, \cite{saber2019} introduced a multi-objective scenario-based planning framework for expanding the BESS, with the goal of minimizing expected wind curtailment and social costs, and \cite{nikoobahkt2020} designed a continuous-time risk-based model for scheduling the BESS on a sub-hourly basis in the day-ahead unit commitment problem. A major disadvantage of these methods is that the BESS serves as a backup in a wind-BESS coupled system, shifting surplus wind generation during low-generation periods to reduce wind curtailment, which negatively impacts the BESS's ability to generate revenue through the electricity market. These methods also require prior knowledge of the wind power uncertainty distribution, which is heavily dependent on historical data.

Bidding strategies that prioritize revenue generation in wind-battery systems have been investigated in several studies. Specifically, \cite{gill2012} introduced a linear programming model to maximize BESS revenue through energy arbitrage, but assumed perfect information on energy prices in the spot market, which is less relevant to real-time bidding. The study in \cite{zhang2018} proposed a market-oriented optimal dispatching strategy for a wind farm equipped with a multi-stage hybrid BESS. Their proposed  dispatch strategy has a priority goal of increasing the market
profits of the wind farm with a costly hydrogen combined cycle energy
storage backup. Studies in \cite{akbari2019,khaloie2020} proposed stochastic optimization models for a wind-battery system participating in both the day-ahead and spinning reserve markets, but bidding simultaneously in electricity markets is more challenging due to the excessive volatility in real-time trading. Furthermore, \cite{nikoobahkt2020,xie2021} developed robust optimization methods to jointly bid in the Spot, Regulation, and Reserve markets, but these optimization-based methods rely heavily on the accuracy of energy price forecasts, which can be difficult to predict due to the substantial market volatility and the complex price drivers~\cite{weron2014}. Moreover, studies in~\cite{urgaokar2011,qin2016,mo2021} proposed online algorithms for BESS scheduling based on Lyapunov optimization~\cite{urgaokar2011,qin2016} and competitive optimization~\cite{mo2021} with performance guarantees. While these methods can be integrated into a co-located wind-battery system if substantial information about system models is available. In contrast, our work seeks a data-driven-based approach that does not require prior knowledge about the system model and can take advantage of tremendous historical data.

In contrast, DRL-based bidding strategies have received relatively less attention in the literature. The research in \cite{yang2020} and \cite{oh2020} introduced multiple variants of the deep Q network (DQN) and the state-action-reward-state-action (SARSA) algorithm, respectively, for a wind farm to participate in the Spot market. Both studies used the BESS as a backup source to mitigate wind forecast errors, without actively participating in the electricity market. However, the algorithmic limitations of DQN and SARSA lead to a discretization of the bidding decision space, resulting in oversimplified bidding strategies. 

To address the deficiencies identified in previous research and bridge the research gap in the development of effective coordination strategies for co-located wind-battery coupled systems, we propose the AggJointDRL method. This approach allows the wind-battery system to dynamically balance the competing objectives of maximizing revenue through participation in the joint market (Spot and Regulation FCAS markets) and minimizing wind curtailment. 

Here, we summarize the novel aspects of our work and differentiate it from existing research as follows: 1) we propose a comprehensive method addressing both wind curtailment management and market bidding for battery energy storage systems, whereas previous research typically focused on only one aspect; 2) our model-free DRL-based bidding strategy for joint-market bidding is novel, as it learns uncertainties associated with wind generation and energy prices from historical observations. This approach does not require prior knowledge or energy price forecasts, a departure from conventional optimization-based methods; 3) our findings underscore the significance of flexible and dynamic coordination between renewable generation and storage in achieving profitability, especially during concurrent bidding in multiple markets. This is a new insight that promotes the coordinated operation between renewable generation and storage. By presenting a practical example of a renewable-storage co-location system, our study holds the potential to support policy-making that encourages renewable-storage asset co-location for wider system benefits. The subsequent section provides a detailed description of the model.

\begin{figure}[!t]
    \centering
    \includegraphics[width=0.9\linewidth]{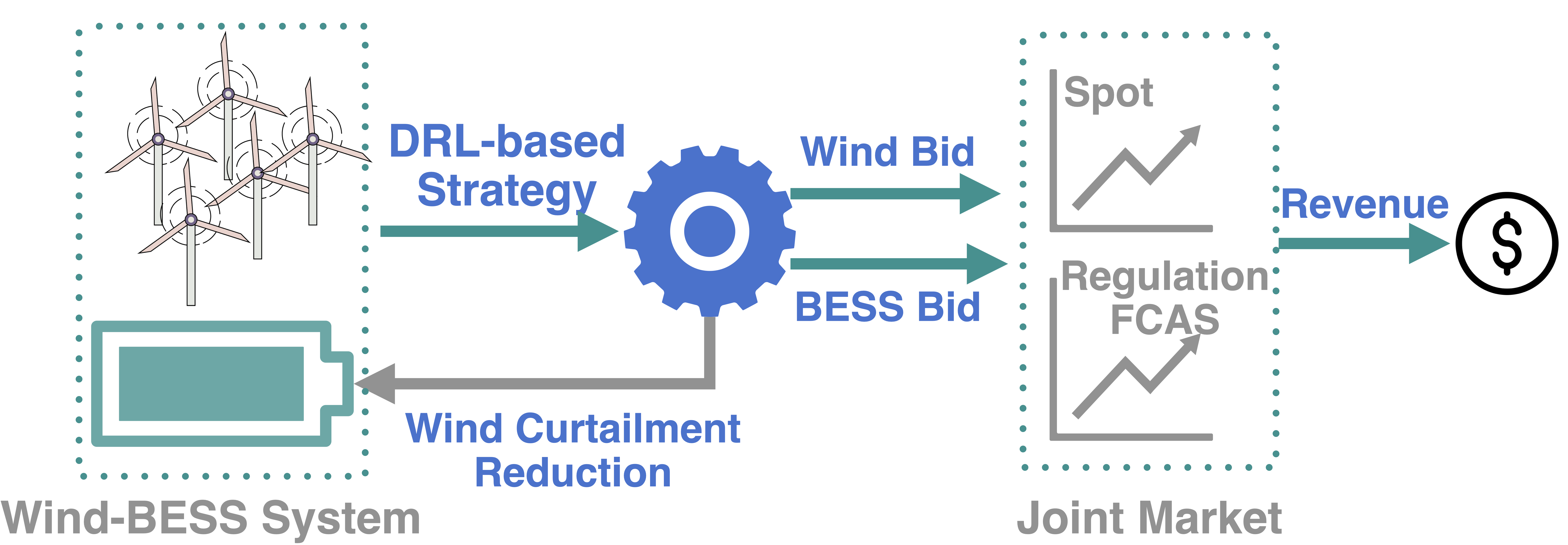}
    \caption{The framework of the AggJointDRL.}
    \label{fig:framework}
\end{figure}

\section{System Model} \label{sec:model}
In developing the AggJointDRL, we assume that the wind-battery system is a price taker, meaning that its bids will not affect the bidding decisions or market clearing outcomes. We also assume that both the co-located wind generators and the BESS are under the ownership and operation of a single system operator. As a result, the BESS acquires onsite wind curtailment at no cost. Furthermore, we assume that there is sufficient capacity within the substation and transmission lines to allow for concurrent export from both facilities. The BESS dynamically manages onsite wind curtailment while engaging in energy arbitrage in the Spot market and providing frequency regulation services in the Regulation FCAS market. The system model is outlined in Fig. \ref{fig:framework}. The context of bidding is discussed in more detail in Section \ref{subsec:model_preliminaries}. Section \ref{subsec:model_revenue} outlines the revenue streams for the wind farm and BESS under various operational conditions, and Section \ref{subsec:model_formulation} formulates the joint-market bidding of the wind-battery system with wind curtailment management.

\subsection{The Australian National Electricity Market} \label{subsec:model_preliminaries}
The Australian National Electricity Market (NEM) comprises approximately 40,000 kilometers of transmission lines and cables that serve around 9 million customers in Australia. It is a wholesale market for the trading of electricity between generators and retailers, connecting the eastern and southern states and territories of the country, and supplying approximately 80\% of all electricity consumption in Australia. The NEM is an energy-only market. The AEMO collects bids from all market participants, forms a bidding stack, and dispatches generators in a least-cost manner~\cite{aemo2021}. In contrast to other electricity markets in the U.S. and Europe, which have an additional day-ahead market, the Australian NEM features only a spot market for wholesale energy trading.

\subsubsection{The Spot market} \label{subsubsec:model_preliminaries_ES}
The Spot market of the NEM serves as a real-time market for the trading of wholesale electricity between generators and loads. The power supply and demand in this market are instantly balanced through a centrally coordinated dispatch process managed by AEMO~\cite{aemo2020}. In the Spot market, generators submit bids, which include both a price and a quantity, every five minutes. AEMO dispatches generators in a cost-effective manner by ranking their bids from low to high to form a bidding stack. This process ensures that power supply-demand imbalances are addressed at the lowest cost possible. The market clearing price, also known as the Spot price, is determined by the generator bids that meet the final demand in the bidding stack~\cite{aemo2017}. If a generator's bid is at or below the Spot price, it will be dispatched at its offered quantity and receive payment at the market clearing price. Often, supply shortages can lead to price spikes in the Spot market. This volatility can create opportunities for market participants to capitalize on fluctuating prices, potentially generating profits from buying low and selling high. Additionally, the volatility may incentivize participants to engage in hedging strategies through various financial instruments, such as contracts for difference, which can lead to more transactions in the wholesale energy market.

\subsubsection{The Regulation FCAS Market} \label{subsubsec:model_preliminaries_regulation}
To maintain stable system frequency and address frequency deviations resulting from the increasing integration of renewables, the FCAS market has been established to procure system services (such as inertia, system restart, and FCAS) from market participants. Similar to the spot market, AEMO arranges MW FCAS offers in a merit order based on costs, where the highest offer that is enabled determines the market clearing price for the respective FCAS category~\cite{aemo2015}. This market can be further divided into Regulation and Contingency FCAS markets~\cite{aemo2015}. In this study, we focus on joint-market bidding in the Spot and Regulation FCAS markets.

Participants registered in the Regulation FCAS market are able to provide Regulation Raise (RR) and Regulation Lower (RL) services to maintain system frequency within the normal operating range of $49.85$Hz to $50.15$Hz. Unlike the direct power exchange in the Spot market, service provision in the Regulation FCAS market is controlled by the automatic generator control (AGC), which sends normalized control signals consecutively to service providers. These signals indicate the desired frequency regulation direction, with positive values representing a frequency ``Raise'' and negative values representing a frequency ``Lower''. For example, if the wind farm or BESS bids to provide RR-FCAS services, it will only output power when receiving frequency-raise regulation signals. We define a series of AGC signals in a single NEM dispatch interval as
\begin{equation}
    \label{eq:def_AGC_signals}
    \bm{s}_t=\left[s_{t,1},\cdots,s_{t,l},\cdots,s_{t,L} \right] \in \left[-\bm{1},\bm{1}\right],
\end{equation}
where $L$ is the length of a series of AGC signals. The absolute value of each AGC signal $|s_{t,l}|$ represents the amount of power that AEMO requires from the bid of service providers to address the instantaneous frequency deviations. The duration of each AGC signal, denoted by $\Delta s$, is $4$ seconds in the NEM. 

\subsection{Revenue of the Wind-Battery System} \label{subsec:model_revenue}
\subsubsection{Wind Farm} \label{subsubsec:model_revenue_wind}
Wind farms in the NEM are typically registered as semi-dispatchable generators and are required to continuously update their forecasted generation availability (derived from onsite wind monitoring devices) to the AEMO. Based on these updates, the market operator issues a dispatch target (in MWh) for the wind farm to meet in the next dispatch interval~\cite{aemo2022}. If the wind farm is able to sell power in both the Spot market and provide regulation raise FCAS (RR-FCAS) services, we introduce a decision variable $v_t^\text{W}\in[0,1]$ to determine the amount of power to be bid in each market based on the dispatch target, i.e., $v_t^\text{W}p_t^\text{W}$ for the Spot market and $(1-v_t^\text{W})p_t^\text{W}$ for the RR-FCAS sub-market.

However, the variable nature of wind power can lead to deviations between the dispatch target and the actual wind generation $p_t^\text{W,Act}$, which in turn affects the amount of power exported by the wind farm. If there is sufficient wind ($p_t^\text{W,Act}>p_t^\text{W}$), we assume that the dispatch target will be fully met. In contrast, if there is a wind shortage caused by forecasting errors or non-compliance with market rules~\cite{aemo2022}, the dispatch target can only be partially met ($p_t^\text{W,Act}<p_t^\text{W}$). We define the actual dispatched wind power as $p_t^\text{W,Dis}=\min\{p_t^\text{W,Act},p_t^\text{W}\}$. The wind farm may also incur a penalty if it fails to meet the dispatch target. The joint-market revenue generated by the wind farm can be expressed as
\begin{equation}
    R^\text{W} = \Delta t \sum_{t=1}^T\left[v_t^\text{W}\rho_t^\text{S}+\left(1-v_t^\text{W}\right)p_t^{\text{RR}}\right]\left(p_t^\text{W,Dis}-\lambda\left|p_t^\text{W,Dis}-p_t^\text{W}\right|\right),
\end{equation}
where $\Delta t$ is the NEM dispatch interval, i.e., $5$ minutes; $T$ is the overall time frame; $\rho_t^\text{S}$ is the Spot price; $\rho_t^\text{RR}$ is the RR-FCAS market clearing price; and $\lambda$ is a penalty coefficient for deviations between the actual wind generation and the AEMO dispatch target~\cite{xie2021}.

\subsubsection{BESS} \label{subsubsec:model_revenue_BESS}
Spot market volatility often motivates the BESS to engage in energy arbitrage. The BESS is also financially rewarded for delivering the RR-FCAS and RL-FCAS services in the Regulation FCAS market. Providing that the BESS cannot operate in both charge and discharge modes concurrently, we introduce two binary variables $v_t^\text{Ch},v_t^\text{Dch}$ to prevent this from happening, which can be formulated as
\begin{equation}
    \label{eq:cons_BESS_ch_dch}
    v_t^\text{Dch} + v_t^\text{Ch} \leq 1, \quad v_t^\text{Dch},v_t^\text{Ch} \in \{0,1\},
\end{equation}
where the BESS sits idle when these two variables are set to zero.

The BESS revenue from the spot and regulation FCAS markets is formulated as
\begin{equation}
    \label{eq:BESS_revenue}
    \begin{aligned}
    R^\text{BESS} = \Delta t \sum_{t=1}^T &\left[\rho_t^\text{S}\left(v_t^\text{Dch}\eta^\text{Dch}p_t^\text{BESS,S} - v_t^\text{Ch}\frac{1}{\eta^\text{Ch}}p_t^\text{BESS,S}\right)\right.\\
    &\left.+\rho_t^\text{RR}v_t^\text{Dch}\eta^\text{Dch}p_t^\text{BESS,Reg}+\rho_t^\text{RL}v_t^\text{Ch}\frac{1}{\eta^\text{Ch}}p_t^\text{BESS,Reg}\right],
    \end{aligned}
\end{equation}
where $\eta^\text{Ch}/\eta^\text{Dch}$ are charge/discharge efficiencies of the BESS; $\rho_t^\text{RR},\rho_t^\text{RL}$ are market clearing prices in the RR-FCAS and RL-FCAS sub-markets; and $p_t^\text{BESS,S}$, $p_t^\text{BESS,Reg}$ are bid power in the spot and regulation FCAS markets.

In addition to purchasing power in the spot market and absorbing excess generation from the grid through the provision of Lower Regulation FCAS services, the BESS can also potentially utilize otherwise curtailed wind generation as a power source for charging. We denote the power intended to be drawn from the onsite wind farm as $p_t^\text{BESS,WC}$. According to the charge/discharge constraint in Equation \eqref{eq:cons_BESS_ch_dch}, the BESS is unable to charge itself using onsite wind curtailment when bidding to discharge in the spot and regulation FCAS markets, which can be described logically as

\begin{equation}
    \label{eq:cons_BESS_dch_wc}
    v_t^\text{Dch}p_t^\text{BESS,WC} = 0.
\end{equation}
The onsite curtailed wind power can be defined as
\begin{equation}
    \label{eq:wind_curtailed_power}
    p_t^\text{W,WC} = \left(p_t^\text{W,Act}-p_t^\text{W}\right)\mathbb{I}\left(p_t^\text{W,Act}>p_t^\text{W}\right),
\end{equation}
where $\mathbb{I}\left(p_t^\text{W,Act}>p_t^\text{W}\right)$ is an indicator of wind curtailment. Thus, the actual power that the BESS draws from the onsite wind farm can be defined as
\begin{equation}
    \label{eq:BESS_wc_response_power}
    \hat{p}_t^\text{BESS,WC} = \min \left\{p_t^\text{BESS,WC},p_t^\text{W,WC}\right\}.
\end{equation}
Moreover, given the fact that the BESS can absorb onsite wind curtailment, the forecasted generation availability of the wind farm $p_t^\text{W}$ can be updated as
\begin{equation}
    \label{eq:corrected_dispatch_target}
    p_t^\text{W} \leftarrow p_t^\text{W} + \hat{p}_t^\text{BESS,WC}.
\end{equation}

Also, frequent charge/discharge leads to cycle aging of the BESS. We define the battery degradation cost as
\begin{equation}
    C^\text{BESS} = c\Delta t\sum_{t=1}^T v_t^\text{Dch}\left(p_t^\text{BESS,S}+p_t^\text{BESS,Reg}\right),
\end{equation}
where we assume only discharge operations cause battery degradation~\cite{bordin2017,anwar2022}, while $c$ is a specific battery technology cost-coefficient in AU\$/MWh~\cite{anwar2022}.

\subsection{Joint-Market Bidding of the Wind-Battery System} \label{subsec:model_formulation}
Considering the distinct revenue streams of the wind farm and the BESS from the spot and regulation FCAS markets, along with the degradation cost of the BESS, we formulate the optimal joint-market bidding of the wind-battery system as an optimization problem whose objective is expressed as
\begin{equation}
    \label{eq:objective}
    R^\text{Joint}=R^\text{W} + R^\text{BESS} - C^\text{BESS}.
\end{equation}

Real-time dispatch of the wind farm and the BESS must be within their installed capacity, which can be formulated as
\begin{align}
    \label{eq:cons_wind_power}
    0&\leq p_t^\text{W}\leq P_\text{max}^\text{W},\\
    \label{eq:cons_BESS_Spot_power}
    0&\leq p_t^\text{BESS,S}\leq P_\text{max}^\text{BESS},\\
    \label{eq:cons_BESS_FCAS_power}
    0&\leq p_t^\text{BESS,Reg}\leq P_\text{max}^\text{BESS},\\
    \label{eq:cons_BESS_WC_power}
    0&\leq p_t^\text{BESS,WC}\leq P_\text{max}^\text{BESS},\\
    \label{eq:cons_BESS_tot_power}
    0&\leq p_t^\text{BESS,S}+p_t^\text{BESS,Reg}+p_t^\text{BESS,WC}\leq P_\text{max}^\text{BESS},
\end{align}
where $P_\text{max}^\text{W}$ and $P_\text{max}^\text{BESS}$ are the installed capacity of the wind farm and the rated power of the BESS, respectively. Eq. \eqref{eq:cons_wind_power} constrains the forecasted availability of the wind farm. Eq. \eqref{eq:cons_BESS_Spot_power} and \eqref{eq:cons_BESS_FCAS_power} constrain the bid power of the BESS in the spot and regulation FCAS markets. Eq. \eqref{eq:cons_BESS_WC_power} represents the power planned to draw from the onsite wind farm must be within the rated power of the BESS. Furthermore, Eq. \eqref{eq:cons_BESS_tot_power} shows that the sum of bids and power drawn from the curtailed wind generation cannot exceed the rated power of the BESS.

Also, charge/discharge operations of the BESS are limited by its current capacity $e_{t-1} + \Delta e_t$, where $e_{t-1}$ is its capacity after the previous dispatch interval, and $\Delta e_t$ is the energy change in the current dispatch interval. The BESS capacity must be within its lower and upper energy limits denoted by $E_\text{min}$ and $E_\text{max}$, which can be formulated as
\begin{equation}
    \label{eq:cons_BESS_energy_change}
    E_\text{min} \leq e_{t-1} + \Delta e_t\leq E_\text{max}.
\end{equation}
The BESS capacity fluctuates due to: 1) power exchange in the spot market; 2) service delivery in the regulation FCAS market; and 3) drawing the otherwise curtailed energy from the onsite wind farm. The energy change from the spot market can be expressed as
\begin{equation}
    \label{eq:BESS_energy_change_Spot}
    \Delta e_t^\text{S} = \Delta t\left(v_t^\text{Ch}-v_t^\text{Dch}\right)p_t^\text{BESS,S}.
\end{equation}

Given that service enablement in the Regulation FCAS market is controlled by the AGC signals, the energy change from the RR-FCAS or RL-FCAS services can be derived by summing energy changes at each AGC signal, which is formulated as
\begin{equation}
    \label{eq:BESS_energy_change_FCAS}
    \begin{aligned}
    \Delta e_t^\text{Reg} = \Delta s & \left[v_t^\text{Ch} \sum_{l=1}^L |s_{t,l}|\mathbb{I}\left(s_{t,l}<0\right) \right.\\
    &\quad \left. - v_t^\text{Dch}\sum_{l=1}^L s_{t,l}\mathbb{I}\left(s_{t,l}\geq 0\right)\right]  p_t^\text{BESS,Reg},
    \end{aligned}
\end{equation}
where $\mathbb{I}(s_{t,l}<0),\mathbb{I}(s_{t,l}\geq 0)$ indicates the frequency-lower and frequency-raise regulation directions, respectively.

The energy change from onsite wind curtailment can be formulated as
\begin{equation}
\label{eq:BESS_energy_change_wc}
\Delta e_t^\text{WC} = \hat{p}_t^\text{BESS,WC} \Delta t,
\end{equation}

Hence, the total energy change of the BESS in the current dispatch interval is written as
\begin{equation}
    \label{eq:BESS_energy_change}
    \Delta e_t = \Delta e_t^\text{S} + \Delta e_t^\text{Reg} + \Delta e_t^\text{WC}.
\end{equation}

The overall formulation of the joint-market bidding optimization is presented as
\begin{equation}
\label{eq:model_formulation}
\begin{aligned}
\max& \quad R^\text{Joint}\\
\text{s.t.}&\quad \text{Eq. \eqref{eq:cons_BESS_ch_dch}, \eqref{eq:cons_BESS_dch_wc}, \eqref{eq:cons_wind_power}--\eqref{eq:cons_BESS_energy_change}}.
\end{aligned}
\end{equation}

\section{Methodology} \label{sec:method}
To maximize the overall revenue of the wind-battery system as defined in Eq. \eqref{eq:model_formulation}, we decompose the continuous bidding problem into two related Markov decision processes (MDPs) for the wind farm and the BESS in Section \ref{subsec:method_MDP}, and then introduce the TD3 algorithm~\cite{fujimoto2018} in Section \ref{subsec:method_TD3} to maximize the expected returns of the resulting MDPs. This facilitates the optimization of the revenue-oriented bidding problem as a whole.

\subsection{MDP Modeling} \label{subsec:method_MDP}
As discussed, multiple factors, such as uncertain wind and energy prices, can influence the bidding behavior of the coupled wind-battery system in the spot and regulation FCAS markets. To better capture the joint-market bidding process, we decompose it into two MDPs (one for the wind farm and one for the BESS), each of which has four elements: the state space $\mathbb{S}^\text{W}$/$\mathbb{S}^\text{BESS}$, the action space $\mathbb{A}^\text{W}$/$\mathbb{A}^\text{BESS}$, the probability space $\mathbb{P}^\text{W}$/$\mathbb{P}^\text{BESS}$, and the reward space $\mathbb{R}^\text{W}$/$\mathbb{R}^\text{BESS}$.

\textbf{State Space $\mathbb{S}$}: All internal (e.g., wind generation and BESS capacity) and external (e.g., energy prices) information of the wind-battery system can be represented as a state $\bm{s}_t$. To guide the BESS's response to wind curtailment, we introduce wind curtailment frequency within the latest $M$ dispatch intervals, denoted by $f_{t}^\text{WC}$, in the state of the BESS. The states of the wind farm and the BESS are defined as
\begin{align}
    \label{eq:MDP_state_wind}
    \bm{s}_t^\text{W} &= \left[ p_{t-1}^\text{W,Act},\rho_{t-1}^\text{S},\rho_{t-1}^\text{RR} \right],\\
    \label{eq:MDP_state_BESS}
    \bm{s}_t^\text{BESS} &= \left[ e_{t-1},f_{t-1}^\text{WC},p_{t-1}^\text{W,Act},\rho_{t-1}^\text{S},\rho_{t-1}^\text{RR},\rho_{t-1}^\text{RL} \right].
\end{align}

\textbf{Action space $\mathbb{A}$}: The actions of the wind farm consist of its forecasted availability $a_t^\text{W}$ (scaled by $P_\text{max}^\text{W}$) and the decision variable $v_t^\text{W}$ (which determines the proportion of bid power in each individual markets). The actions of the BESS include the power bid in the joint market ($a_t^\text{BESS,S},a_t^\text{BESS,Reg}$) and drawn from the onsite wind farm ($a_t^\text{BESS,WC}$), which are all scaled by $P_\text{max}^\text{BESS}$. The actions of the BESS also include the charge/discharge variables $v_t^\text{Ch}/v_t^\text{Dch}$. We define actions of the wind farm and the BESS as
\begin{align}
    \label{eq:MDP_action_wind}
    \bm{a}_t^\text{W} &= \left[ a_t^\text{W},v_t^\text{W} \right],\\
    \label{eq:MDP_action_BESS}
    \bm{a}_t^\text{BESS} &= \left[ v_t^\text{Dch},v_t^\text{Ch},a_t^\text{BESS,S},a_t^\text{BESS,Reg},a_t^\text{BESS,WC} \right].
\end{align}

\textbf{Probability space $\mathbb{P}$}: The probability space refers to the set of probabilities for transitioning to the next state after taking a deterministic action, which is defined as $\mathbb{P}\left(\bm{s}_{t+1}|\bm{s}_t,\bm{a}_t\right)$.

\textbf{Reward Space $\mathbb{R}$}: The wind farm and the BESS receive rewards after taking action $\bm{a}_t$ at state $\bm{s}_t$, which reflect the effectiveness of the bidding decision. To maximize the overall revenue, it is essential to design appropriate reward functions for both the wind farm and the BESS to make better bidding decisions.

To account for wind generation uncertainty, update accurate dispatch targets, and take advantage of the flexibility of the joint market, we formulate the reward function of the wind farm as
\begin{equation}
    \label{eq:reward_wind}
    r_t^\text{W} = \left[v_t^\text{W}\rho_t^\text{S} + \left(1-v_t^\text{W}\right)\rho_t^\text{RR}\right] \left(\min\left\{a_t^\text{W},a_t^\text{W,Act}\right\}-\lambda \left|a_t^\text{W}-a_t^\text{W,Act}\right|\right),
\end{equation}
with the normalized actual wind generation $a_t^\text{W,Act}$ defined as
\begin{equation}
    \label{eq:normalized_actual_wind_power}
    a_t^\text{W,Act} = \frac{p_t^\text{W,Act}}{P_\text{max}^\text{W}}.
\end{equation}

Effective BESS energy arbitrage in the spot market is enabled by the introduction of two charge/discharge indicators, denoted by $\mathbb{I}_t^\text{Ch}$/$\mathbb{I}_t^\text{Dch}$, formulated as
\begin{align}
    \label{eq:ch_indicator}
    \mathbb{I}_t^\text{Ch} &= \text{sgn}\left(\bar{\rho}_t^\text{S}-\rho_t^\text{S}\right),\\
    \label{eq:dch_indicator}
    \mathbb{I}_t^\text{Dch} &= \text{sgn}\left(\rho_t^\text{S}-\bar{\rho}_t^\text{S}\right),
\end{align}
where $\text{sgn}(\cdot)$ is the sign function and $\bar{\rho}_t^\text{S}$ is the exponential moving average of the spot price, which is defined as
\begin{equation}
    \label{eq:moving_avg_Spot_price}
    \bar{\rho}_t^\text{S} = \tau^\text{S}\bar{\rho}_{t-1}^\text{S} + \left(1-\tau^\text{S}\right)\rho_t^\text{S},
\end{equation}
where $\tau^\text{S}$ is a smoothing parameter. The charge/discharge indicators incentivize the BESS to buy low ($\rho_t^\text{S}<\bar{\rho}_t^\text{S}$) and sell high ($\rho_t^\text{S}>\bar{\rho}_t^\text{S}$). Any bids violating such a guideline will be penalized. Hence, the arbitrage rewards can be formulated as
\begin{equation}
    \label{eq:BESS_Spot_reward}
    r_t^\text{BESS,S} = a_t^\text{BESS,S} |\rho_t^\text{S} - \bar{\rho}_t^\text{S}| \left(\mathbb{I}_t^\text{Ch}v_t^\text{Ch}\frac{1}{\eta^\text{Ch}} + \mathbb{I}_t^\text{Dch}v_t^\text{Dch}\eta^\text{Dch}\right).
\end{equation}

Moreover, rewards from the Regulation FCAS market are formulated as
\begin{equation}
    \label{eq:BESS_FCAS_reward}
    r_t^\text{BESS,Reg} = a_t^\text{BESS,Reg}\left(v_t^\text{Ch}\frac{1}{\eta^\text{Ch}}\rho_t^\text{RL}+v_t^\text{Dch}\eta^\text{Dch}\rho_t^\text{RR}\right).
\end{equation}

Also, the BESS receives positive rewards when it reduces onsite wind curtailment, which can be formulated as 
\begin{equation}
    \label{eq:BESS_wc_reward}
    \begin{aligned}
    r_t^\text{BESS,WC} = \lambda & \left[v_t^\text{W}\rho_t^\text{S}+\left(1-v_t^\text{W}\right)\rho_t^\text{RR}\right] \\
    &\times \min\left\{a_t^\text{BESS,WC},a_t^\text{W,WC}\right\} f_t^\text{WC}\frac{1}{\eta^\text{Ch}},
    \end{aligned}
\end{equation}
% \begin{equation}
%     \label{eq:BESS_wc_reward}
%     r_t^\text{BESS,WC} = \lambda\left[v_t^\text{W}\rho_t^\text{S}+\left(1-v_t^\text{W}\right)\rho_t^\text{RR}\right]  \min\left\{a_t^\text{BESS,WC},a_t^\text{W,WC}\right\} f_t^\text{WC}\frac{1}{\eta^\text{Ch}},
% \end{equation}
with the normalized onsite curtailed wind power defined as
\begin{equation}
    \label{eq:normalized_curtailed_wind_power}
    a_t^\text{W,WC} = \frac{p_t^\text{W,WC}}{P_\text{max}^\text{W}}.
\end{equation}

The BESS reward function combines the bidding rewards and wind curtailment mitigation rewards, and is defined as
\begin{equation}
    \label{eq_BESS_tot_reward}
    r^\text{BESS}_t = r_t^\text{BESS,S} + r_t^\text{BESS,Reg} + r_t^\text{BESS,WC}.
\end{equation}

\subsection{Learning Optimal Bidding Strategy via TD3}
\label{subsec:method_TD3}
We introduce a state-of-the-art off-policy DRL algorithm, referred to as TD3~\cite{fujimoto2018}, to optimize the derived MDPs (where the same TD3 structure is adopted). For brevity, we present the details of the TD3 without specifying notations of the MDPs mentioned earlier. TD3 aims to learn an optimal action strategy, denoted by $\pi(\bm{a}_t|\bm{s}_t)$, that maximizes the expected returns over a finite horizon, which can be formulated as
\begin{equation}
    \label{eq:DRL_objective}
    J_\pi = \mathbb{E}_{\bm{s}_t\sim \mathbb{P},\bm{a}_t\sim \pi(\bm{s}_t)} \left[ R_1 \right],
\end{equation}
with the discounted expected return defined as
\begin{equation}
    \label{eq:expected_return}
    R_t = \sum_{t'=t}^T \gamma^{t'-t}r_{t'},
\end{equation}
where $\gamma$ is the discounted factor.

TD3 is facilitated with two essential functions following an actor-critic framework, where the actor function $\pi(\bm{a}_t|\bm{s}_t)$ determines an action for state transition and the critic function (also known as the Q function) $Q(\bm{a}_t,\bm{s}_t)$ reflects the effectiveness of the state-action pair. The Q function can be formulated as
\begin{equation}
    \label{eq:Q_func}
    Q\left(\bm{a}_t,\bm{s}_t\right) = \mathbb{E}_{\bm{s}_t\sim \mathbb{P},\bm{a}_t\sim\pi(\bm{s}_t)} \left[R_t|\bm{s}_t,\bm{a}_t\right].
\end{equation}
With the help of the Q function, the objective of the actor function in Eq. \eqref{eq:DRL_objective} can be rewritten as
\begin{equation}
    \label{eq:DRL_objective_rewritten}
    J_\pi = \mathbb{E}_{\bm{s}\sim\mathbb{P}}\left[ Q(\bm{s}_1,\pi(\bm{s}_1)) \right].
\end{equation}

We adopt neural networks as function approximators for estimating the actor function $\pi_\phi$ and the Q function $Q_\theta$, where $\phi,\theta$ represent the corresponding neural network parameters. The Adam optimizer~\cite{goodfellow2016} is used to perform gradient descent for updating these neural networks.

\textbf{Update $\pi_\theta$}: The actor network is updated by maximizing its objective defined in Eq. \eqref{eq:DRL_objective_rewritten}, whose gradient can be formulated as
\begin{equation}
    \label{eq:gradient_pi}
    \nabla_\phi J_\pi(\phi) = \nabla_\phi \mathbb{E}_{\bm{s}\sim \mathbb{B}} \left[ Q_\theta\left( \bm{s}_1,\pi_\phi(\bm{s}_1) \right) \right],
\end{equation}
where the replay buffer $\mathbb{B}$ is introduced to store transition information of the MDP. One transition is defined as $\{\bm{s}_t,\bm{a}_t,r_t,\bm{s}_{t+1}\}$. 

In particular, to ensure that the sum of bids and the power drawn from the onsite wind farm does not exceed the installed capacity of the BESS as defined in Eq. \eqref{eq:cons_BESS_tot_power}, we propose an associated loss function within the TD3 of the BESS, which is formulated as
\begin{equation}
    \label{eq:loss_func}
    L_\pi(\phi) = a_t^\text{BESS,sum} \mathbb{I}\left(a_t^\text{BESS,sum}>1\right),
\end{equation}
with the sum of BESS actions defined as
\begin{equation}
    \label{eq:BESS_sum_power}
    a_t^\text{BESS,sum} = a_t^\text{BESS,S} + a_t^\text{BESS,Reg} + a_t^\text{BESS,WC}.
\end{equation}

Thus, the gradient of the BESS actor network can be rewritten as
\begin{equation}
    \label{eq:graident_pi_BESS}
    \nabla_\phi J_\pi(\phi) \leftarrow \nabla_\phi J_\pi(\phi) - \beta^L \nabla_\phi \mathbb{E}_{\bm{a}\sim\mathbb{B}}\left[ L_\pi(\phi) \right],
\end{equation}
where $\beta^L$ is the coefficient of the proposed loss function. The general gradient descent process of the actor network is formulated as
\begin{equation}
    \label{eq:gradient_descent_pi}
    \phi \leftarrow \phi - \eta^\phi \nabla_\phi \left[-J_\pi(\phi)\right],
\end{equation}
where $\eta^\phi$ is the learning rate of the actor network. Moreover, delaying the update of the actor network can effectively reduce the estimation variance of the Q function in TD3~\cite{fujimoto2018}, where a delaying factor $d$ is introduced to control the asynchronous update between the actor and Q networks.

\begin{figure}[!t]
    \centering
    \includegraphics[width=0.9\linewidth]{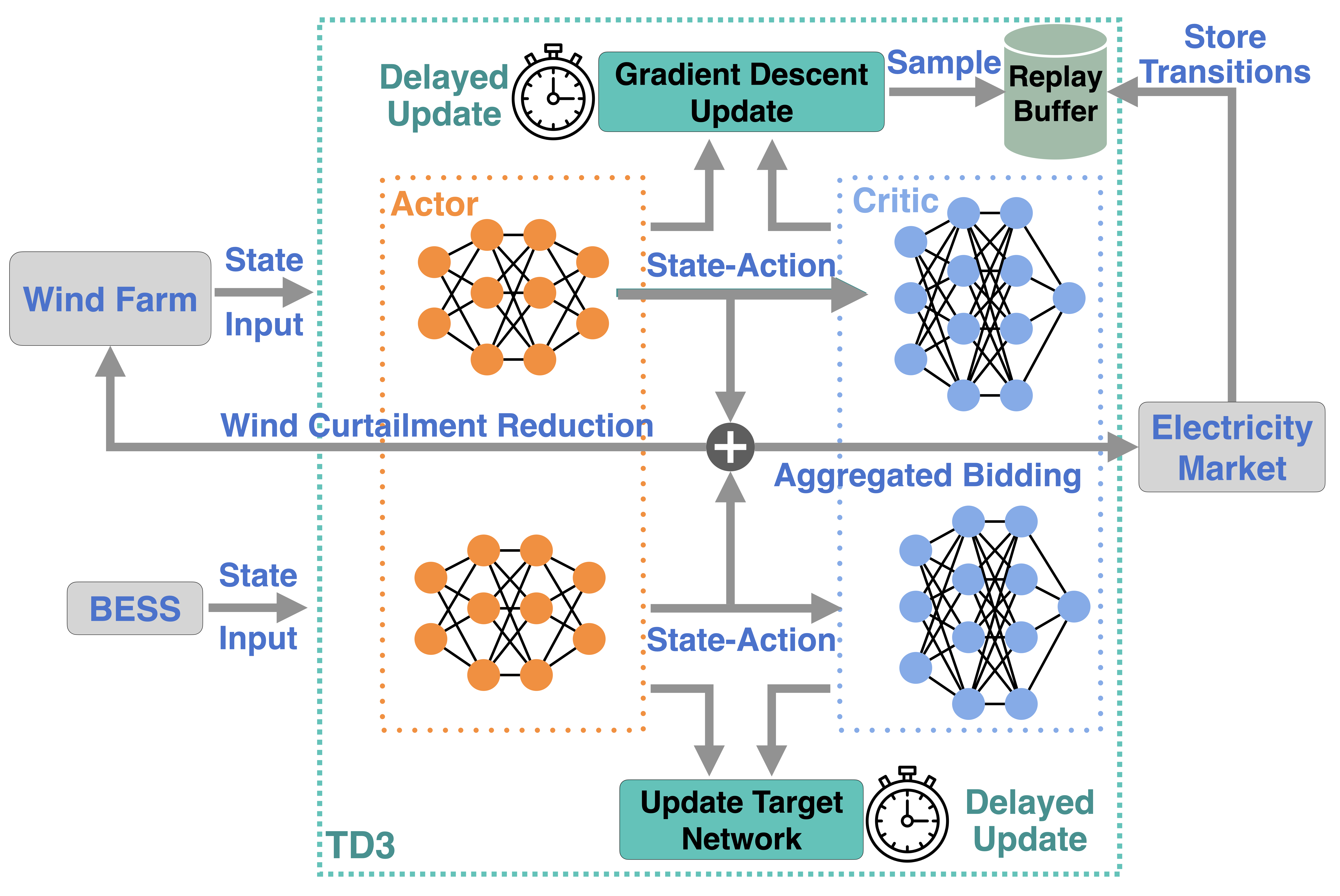}
    \caption{The structure of the TD3-empowered AggJointDRL.}
    \label{fig:TD3_structure}
\end{figure}

\textbf{Update $Q_\theta$}: The Q network is updated by minimizing the residual error of the Bellman equation. The Bellman equation~\cite{sutton2018} is defined as
\begin{equation}
    \label{eq:Bellman_eq}
    Q(\bm{s}_t,\bm{a}_t) = r_t + \gamma \mathbb{E}_{\bm{s}_{t+1}\sim\mathbb{P},\bm{a}_{t+1}\sim\pi(\bm{s}_{t+1})}\left[Q\left(\bm{s}_{t+1},\bm{a}_{t+1}\right)\right],
\end{equation}
where the left-hand side of the equation is estimated by the Q network $Q_\theta$ and the right-hand side is by a parameter-frozen target Q network $Q_{\hat{\theta}}$ ($\hat{\theta}$ is the parameter of the target Q network). Hence, the objective of the Q network can be formulated as
\begin{equation}
    \label{eq:objective_Q_net}
    J_Q(\theta) = \mathbb{E}_{\bm{s}_t\sim\mathbb{B}}\left[\frac{1}{2}\left[Q_\theta\left(\bm{s}_t,\pi_\phi(\bm{s}_t)\right)-Q_{\hat{\theta}}\left(\bm{s}_t,\pi_{\hat{\phi}}(\bm{s}_t)\right)\right]^2\right],
\end{equation}
where a target actor network $\pi_{\hat{\phi}}$ is introduced to stabilize the update process and assist the target Q network in estimating the Q function. 

Similar to the actor network, the gradient descent process of the Q network can be formulated as
\begin{equation}
    \label{eq:gradient_descent_Q_net}
    \theta \leftarrow \theta - \eta^\theta \nabla_\theta J_Q(\theta),
\end{equation}
where $\eta^\theta$ is the learning rate of the Q network as well.

In addressing the issues of overestimation~\cite{hasselt2010} in the Bellman equation, a clipped double Q-learning technique has been introduced in TD3, where one more Q network, denoted by $Q_{\theta_2}$ (noting that we denoted the aforementioned Q network $Q_\theta$ by $Q_{\theta_1}$), is adopted to minimize the estimation bias. Therefore, the estimation of Q function via the target networks, i.e., $Q_{\hat{\theta}}$ and $\pi_{\hat{\phi}}$, can be reformulated as
\begin{equation}
    \label{eq:Q_func_target_net}
    Q_{\hat{\theta}}\left(\bm{s}_t,\pi_{\hat{\phi}}(\bm{s}_t)\right) = r_t + \gamma \min_{i=\{1,2\}} \mathbb{E}_{\bm{s}_{t+1}\sim\mathbb{P}}\left[Q_{\hat{\theta}_i}\left(\bm{s}_{t+1},\pi_{\hat{\phi}}(\bm{s}_{t+1})\right)\right],
\end{equation}
where $\hat{\theta}_1,\hat{\theta}_2$ are target Q network parameters.

\textbf{Update target networks}: In the TD3 algorithm, all target networks, including the target actor network $\pi_{\hat{\phi}}$ and the two target Q networks $Q_{\hat{\theta}_1},Q_{\hat{\theta}_2}$, are updated in a moving-average manner  in synchrony with the actor network, which can be formulated as
\begin{align}
    \label{eq:update_target_actor_net}
    \hat{\phi}&\leftarrow\tau^\phi\phi+(1-\tau^\phi)\hat{\phi},\\
    \label{eq:update_target_Q_net}
    \hat{\theta}_i&\leftarrow\tau^\theta\theta_i+(1-\tau^\theta)\hat{\theta}_i,\hspace{0.25em} i=\{1,2\},
\end{align}
where $\tau^\phi,\tau^\theta$ are smoothing parameters.  

To summarize, the overall structure of our TD3-empowered AggJointDRL is illustrated in Fig. \ref{fig:TD3_structure}, with a detailed algorithmic procedure provided in Algorithm \ref{algo:bidding_strategy}.

\begin{algorithm}[!t]
\caption{The TD3-empowered AggJointDRL.}
\label{algo:bidding_strategy}
\begin{algorithmic}
\STATE Initialize actor and Q networks of both the wind farm and BESS $\phi^\text{W},\theta^\text{W}_1,\theta^\text{W}_2,\phi^\text{BESS},\theta^\text{BESS}_1,\theta^\text{BESS}_2$.
\STATE Initialize target networks with their corresponding network parameters.
\STATE Initialize the replay buffers $\mathbb{B}^\text{W},\mathbb{B}^\text{BESS}$.
\FOR{$t=1,\cdots,T$}
\STATE Prepare the current states $\bm{s}_t^\text{W},\bm{s}_t^\text{BESS}$.
\STATE Get actions $\bm{a}_t^\text{W},\bm{a}_t^\text{BESS}$ through action networks and receive associated rewards $r_t^\text{W},r_t^\text{BESS}$.
\IF{the BESS bids violate its energy limits}
\STATE $\bm{a}_t^\text{BESS} \leftarrow \bm{0}$.
\ENDIF
\STATE Transit into the next states $\bm{s}_{t+1}^\text{W},\bm{s}_{t+1}^\text{BESS}$ via $\mathbb{P}^\text{W}$ and $\mathbb{P}^\text{BESS}$.
\STATE Store transitions $\{\bm{s}_t^\text{W},\bm{a}_t^\text{W},r_t^\text{W},\bm{s}_{t+1}^\text{W}\},\{\bm{s}_t^\text{BESS},\bm{a}_t^\text{B},r_t^\text{BESS},\bm{s}_{t+1}^\text{BESS}\}$ into replay buffers $\mathbb{B}^\text{W}$ and $\mathbb{B}^\text{BESS}$.
\STATE Update Q networks via gradient descent.
\IF{$t\mod d=0$}
\STATE Update actor and target networks.
\ENDIF
\ENDFOR
\end{algorithmic}
\end{algorithm}

\begin{table}[!t]
    \centering
    % \vspace*{-6mm}
    \caption{The initialized parameters.}
    \begin{tabular}{cc||cc}
    \hline
    $\Delta s$ & $4$ secs & $L$ & $75$\\
    $\Delta t$ & $5$ mins & $\lambda$ & $1.5$\\
    $\eta^\text{Dch},\eta^\text{Ch}$ & $0.95$ & $c$ & $1$ AU\$/MWh\\
    $P_\text{max}^\text{W}$ & $67$ MW & $P_\text{max}^\text{BESS}$ & $10$ MW\\
    $E_\text{min}$ & $0.5$ MWh & $E_\text{max}$ & $9.5$ MWh\\
    $M$ & $10$ & $\tau^\text{S}$ & $0.9$\\
    $\gamma$ & $0.99$ & $\beta^L$ & $10$\\
    $\eta^\phi,\eta^\theta$ & $0.0003$ & $\tau^\phi,\tau^\theta$ & $0.01$\\
    \hline
    \end{tabular}
    \label{tab:parameters}
    % \vspace*{-2mm}
\end{table}

\begin{figure}[!t]
    \centering
    \includegraphics[width=0.6\linewidth]{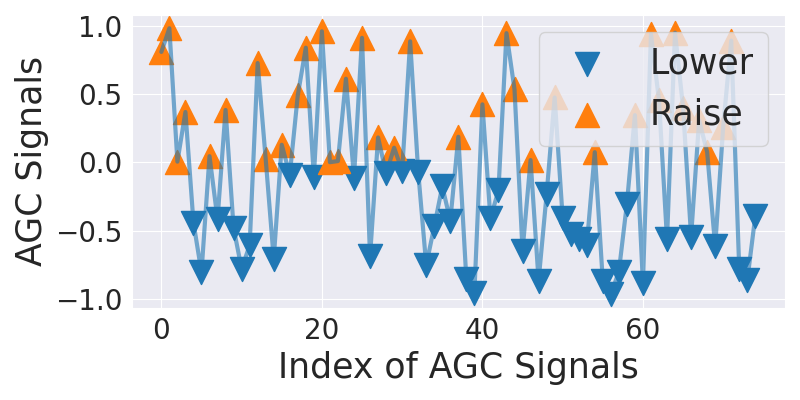}
    \caption{The AGC signals in one dispatch interval. Negative/positive signal values represent the frequency-lower and frequency-raise regulation directions.}
    \label{fig:AGC_signal}
\end{figure}

\section{Experiments and Results} \label{sec:exp}
\subsection{Experimental Settings} \label{subsec:exp_setting}
The wind generation data used in this study were collected from the Oaklands Hill Wind Farm in Victoria -- one of the five jurisdictions of the National Electricity Market in Australia. We used Victoria spot prices from 2018~\cite{aemo_data} to train and evaluate our AggJointDRL method, with the first eleven months of data used for training and the final month used for evaluation. The battery energy storage system had a storage capacity of 10 MWh, with a minimum allowable state of charge of 5\% and a maximum allowable state of charge of 95\%. We utilized an Nvidia TITAN RTX graphics processing unit for training the algorithm, and the initialized parameters are listed in Table \ref{tab:parameters}. We compared three scenarios in which the wind-battery system could bid: 1) the spot market only; 2) the regulation FCAS market only; and 3) the joint spot-regulation FCAS markets. To simulate frequency regulation directions (e.g., lower or raise), we synthesized automatic generation control (AGC) signals conforming to a uniform distribution $U\sim(-1,1)$ for both the wind farm and the BESS when participating in the Regulation FCAS market. An example of AGC signals in one dispatch interval is shown in Figure \ref{fig:AGC_signal}. Though we conduct the simulations in the Australian National Electricity Market serving as a case study, we emphasize that our proposed bidding strategy for the co-located wind-battery system is adaptable and applicable to other electricity markets, for example markets in different regions of the U.S. and Europe.

\begin{table*}[!t]
    \centering
    % \vspace*{-6mm}
    \caption{The evaluation revenue (in AU\$) and time costs (in minutes) of the P\&O benchmark and our AggJointDRL in all three bidding scenarios.}
    \begin{tabular}{|c|c|c|c|c|c|}
    \cline{3-6}
    \multicolumn{2}{c|}{} & Strategy & Spot & Reg FCAS & Joint\\
    \hline 
    \multirow{9}{*}{Revenue} & \multirow{3}{*}{Wind} & P\&O & $829,604$ & $451,156$ & $889,672$\\
    \cline{3-6}
    & & \textbf{Ours} & $979,784$ & $540,690$ & $1,098,686$\\
    \cline{3-6}
    & & Boost & $\bm{18\%}$ & $\bm{20\%}$ & $\bm{23\%}$\\
    \cline{2-6}
    & \multirow{3}{*}{BESS} & P\&O & $62,444$ & $178,643$ & $412,363$\\
    \cline{3-6}
    & & \textbf{Ours} & $113,292$ & $247,759$ & $524,917$\\
    \cline{3-6}
    & & Boost & $\bm{81\%}$ & $\bm{39\%}$ & $\bm{27\%}$\\
    \cline{2-6}
    & \multirow{3}{*}{Total} & P\&O & $892,048$ & $629,799$ & $1,302,035$\\
    \cline{3-6}
    & & \textbf{Ours} & $1,093,076$ & $788,449$ & $1,623,603$\\
    \cline{3-6}
    & & Boost & $\bm{23\%}$ & $\bm{25\%}$ & $\bm{25\%}$\\
    \hline
    \multirow{6}{*}{Time Cost} & \multirow{2}{*}{Train} & P\&O & $1.8$ & $3.4$ & $3.7$\\
    \cline{3-6}
    & & Ours & $92.4$ & $89.7$ & $91.9$\\
    \cline{2-6}
    & \multirow{2}{*}{Evaluate} & P\&O & $36.3$ & $71.2$ & $94.6$\\
    \cline{3-6}
    & & \textbf{Ours} & $\bm{0.1}$ & $\bm{0.1}$ & $\bm{0.1}$\\
    \cline{2-6}
    & \multirow{2}{*}{Total} & P\&O & $38.1$ & $74.6$ & $98.3$\\
    \cline{3-6}
    & & Ours & $92.5$ & $89.8$ & $\bm{92.0}$\\
    \hline
    \end{tabular}
    \label{tab:evaluation_results_benchmark_comp}
    % \vspace*{-2mm}
\end{table*}

\subsection{Effectiveness of the AggJointDRL} \label{subsec:exp_effectiveness_AggJointDRL}

\begin{figure}[!t]
    \centering
    \includegraphics[width=\linewidth]{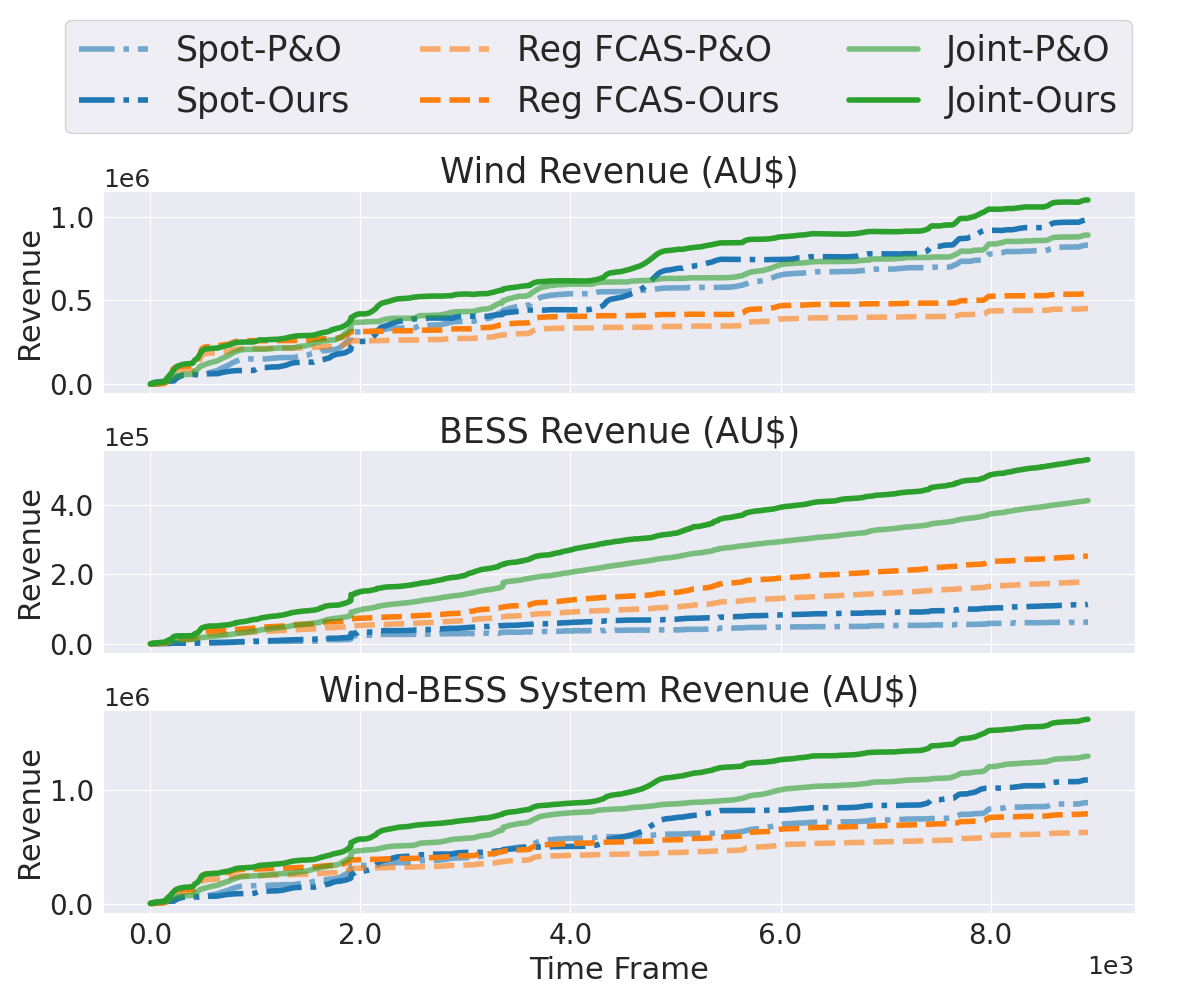}
    \caption{The evaluation revenue of the P\&O benchmark and our AggJointDRL in all three bidding scenarios.}
    \label{fig:evaluation_results_benchmark_comp}
\end{figure}

\subsubsection{Benchmark Comparison} \label{subsubsec:exp_effectiveness_AggJointDRL_benchmark_comp}
To examine the effectiveness of our AggJointDRL method, we conducted a comparison using a predict-and-optimize (P\&O) benchmark. The P\&O method applies long short-term memory (LSTM) networks for forecasting both wind availability and energy prices, and employs a mixed integer linear programming solver from the PuLP library~\cite{mitchell2011} to solve the revenue maximization bidding problem. The results of this comparison, including revenue generated from both methods in three different bidding scenarios, are depicted in Figure \ref{fig:evaluation_results_benchmark_comp} and summarized in Table \ref{tab:evaluation_results_benchmark_comp} for cross comparison.

\begin{figure}[!t]
    \centering
    \includegraphics[width=0.7\linewidth]{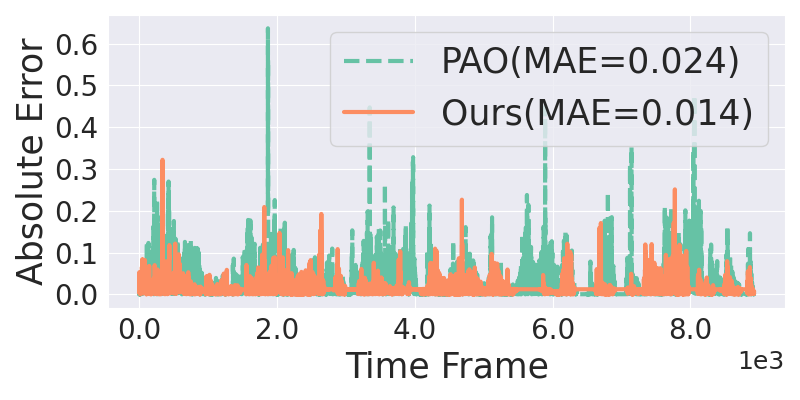}
    \caption{The absolute errors between the dispatch targets and the actual wind generation.}
    \label{fig:evaluation_results_wind_abs_error}
\end{figure}

The simulation results indicate that our AggJointDRL method outperformed the P\&O benchmark in all three bidding scenarios by 23\%, 25\%, and 25\%, respectively. Notably, Table \ref{tab:evaluation_results_benchmark_comp} shows that our method significantly improved BESS bidding performance in the spot market by 81\%, in the regulation FCAS market by 39\%, and in the joint market by 27\%. Additionally, revenue generated by the wind farm using our method was significantly higher than that of the benchmark. Figure \ref{fig:evaluation_results_wind_abs_error} illustrates the absolute errors between forecasted wind availability and actual wind generation when the wind farm participated in the spot market. This figure demonstrates that the AggJointDRL has a superior ability to capture the uncertainty of wind generation, as evidenced by its lower mean absolute error (MAE), which allows the wind farm to more accurately update its dispatch target and avoid financial penalties for dispatch deviations.

Although the AggJointDRL requires a longer training time, as shown in Table \ref{tab:evaluation_results_benchmark_comp}, it can make bidding decisions at a significantly faster rate, with a decision time of only 10 seconds for one-month bidding. In contrast, the evaluation time cost of the P\&O method significantly increases with the number of participated markets. Thus, a well-trained AggJointDRL is better suited for real-time bidding, where accurate and rapid decision-making is essential.

\begin{figure}[!t]
    \centering
    \includegraphics[width=0.7\linewidth]{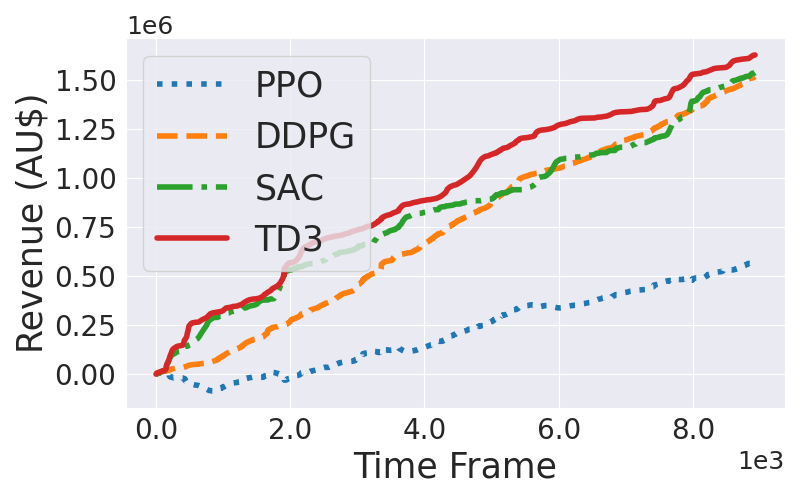}
    \caption{The revenue generated by the wind-BESS system through joint-market bidding using various algorithms.}
    \label{fig:evaluation_results_DRL_comp}
\end{figure}

In addition, we trained our AggJointDRL using a variety of different deep reinforcement learning (DRL) algorithms, including proximal policy optimization (PPO)\cite{schulman2017}, deep deterministic policy gradient (DDPG)\cite{lillicrap2016}, and soft actor critic (SAC)~\cite{haarnoja2018}. The results of this comparison are shown in Figure \ref{fig:evaluation_results_DRL_comp}. The bidding performance of the three off-policy algorithms (DDPG, SAC, and TD3) was similar to each other and significantly outperformed the on-policy algorithm PPO. This may be due to the fact that the on-policy algorithm does not utilize a replay buffer to store all historical observations, resulting in less exploitation of the state spaces (i.e., $\mathbb{S}^\text{W}$,$\mathbb{S}^\text{BESS}$) of both the wind farm and the BESS. These evaluation results suggest that our proposed joint-market bidding framework may be compatible with various off-policy DRL algorithms.

\subsubsection{Profitability of the Regulation FCAS Market} \label{subsubsec:exp_effectiveness_AggJointDRL_FCAS_profitability}
Table \ref{tab:evaluation_results_benchmark_comp} shows that the BESS generates higher revenue when participating in the Regulation FCAS market compared to the spot market. This is because, once registered in the Regulation FCAS market, the BESS can receive rewards for providing both RR-FCAS and RL-FCAS services through capacity reserve rather than actual service delivery. In contrast, in the spot market, the BESS must purchase power to charge itself, which can result in revenue losses (when the spot price is non-negative) and contribute to the revenue gap between these two bidding scenarios. 

Table \ref{tab:regulation_FCAS_price_annual_summary} provides statistics on the historical market clearing prices in the Regulation FCAS market from 2016 to 2021. The rapid adoption of renewable energy sources has introduced uncertainty and disturbance to the power grid, leading to increased procurement of system services (particularly FCAS services) from market participants by the energy market operator. This trend has contributed to the growth and development of the Regulation FCAS market, along with the rising average market clearing prices and increased market volatility, as reflected in the higher standard deviations observed annually in Table \ref{tab:regulation_FCAS_price_annual_summary}. The observed market volatility also highlights the underlying economic opportunity in the Regulation FCAS market, incentivizing electricity market participants (especially lithium-iron batteries than can respond within a second) to deliver Regulation FCAS services rather than engage in arbitrage in the spot market.

\begin{table*}[!t]
    \centering
    % \vspace*{-6mm}
    \caption{Statistics of historical Regulation FCAS market clearing prices.}
    \begin{tabular}{|c|c|c|c|c|c|c|c|}
    \cline{3-8}
    \multicolumn{2}{c|}{} & 2016 & 2017 & 2018 & 2019 & 2020 & 2021\\
    \hline
    \multirow{2}{*}{RR-FCAS} & mean (AU\$/MWs) & $13$ & $30$ & $24$ & $39$ & $22$ & $19$\\
    \cline{2-8}
    & std (AU\$/MWs) & $14$ & $31$ & $68$ & $62$ & $192$ & $110$\\
    \hline
    \multirow{2}{*}{RL-FCAS} & mean (AU\$/MWs) & $6$ & $24$ & $11$ & $19$ & $10$ & $13$\\
    \cline{2-8}
    & std (AU\$MWs) & $6$ & $16$ & $9$ & $11$ & $7$ & $54$\\
    \hline
    \end{tabular}
    \label{tab:regulation_FCAS_price_annual_summary}
    % \vspace*{-2mm}
\end{table*}

\subsubsection{Effectiveness of Joint-Market Bidding} \label{subsubsec:exp_effectiveness_AggJointDRL_joint_market}
As previously mentioned, the Regulation FCAS market presents a significant opportunity for profitability. By participating jointly in both the spot and regulation FCAS markets, it is possible to fully exploit the flexibility of both markets, resulting in significantly higher revenue compared to individual market participation, as demonstrated in Table \ref{tab:evaluation_results_benchmark_comp}. It is worth noting that the revenue generated from the BESS through joint market participation even exceeds the combined revenue from engaging in the individual markets.

\begin{figure}[!t]
    \centering
    \includegraphics[width=0.7\linewidth]{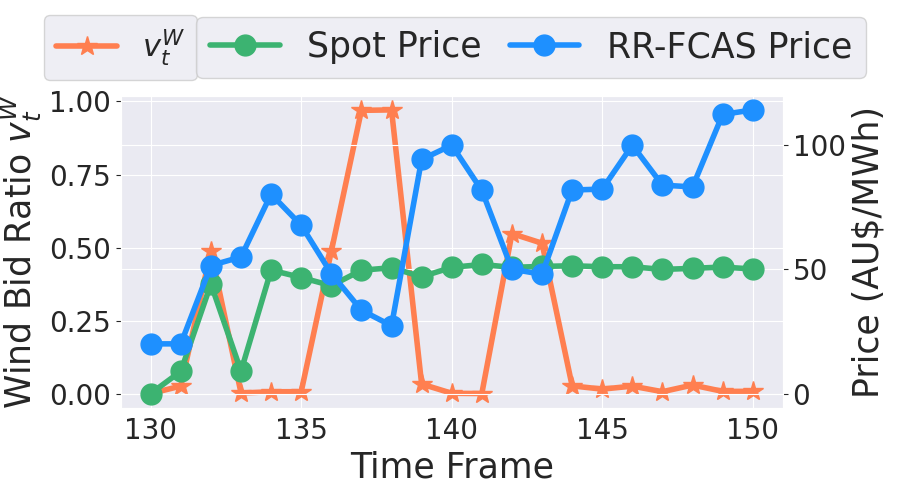}
    \caption{The wind farm bidding patterns in the joint market.}
    \label{fig:case_study_wind_joint_market}
\end{figure}

Bidding in the Regulation FCAS market can generate higher revenue if the FCAS market clearing prices are higher than the spot price in the spot market, and vice versa. Figure \ref{fig:case_study_wind_joint_market} provides an example of how the wind farm can take advantage of the volatility in both markets in the joint market scenario. In this case, the wind farm tends to bid on a greater amount of power (i.e., relatively low values of the decision variable $v_t^\text{W}$) in the RR-FCAS sub-market when the RR-FCAS market clearing price is higher than the spot price. This demonstrates the benefits of participating in both markets jointly, as it allows the wind farm to maximize profits by capitalizing on fluctuations in prices from both markets.

\begin{table}[!t]
    \centering
    % \vspace*{-6mm}
    \caption{The wind curtailment responses (in MWh) in the P\&O and AggJointDRL.}
    \begin{tabular}{c|c|c|c|}
    \cline{2-4}
    & Spot & Reg. FCAS & Joint\\
    \hline
    P\&O & $130/623$ & $53/623$ & $78/623$\\
    % MW 1568/7474, 632/7474, 936/7474
    \hline
    \textbf{AggJointDRL} & $313/437$ & $235/371$ & $257/415$\\
    % MW 3756/5244, 2817/4457, 3080/4982
    \hline
    \end{tabular}
    \label{tab:wc_responses_benchmark_comp}
    % \vspace*{-2mm}
\end{table}

\subsection{Wind Curtailment Management} \label{subsec:exp_wc_reduction}
\begin{figure}[!t]
    \centering
    \includegraphics[width=0.8\linewidth]{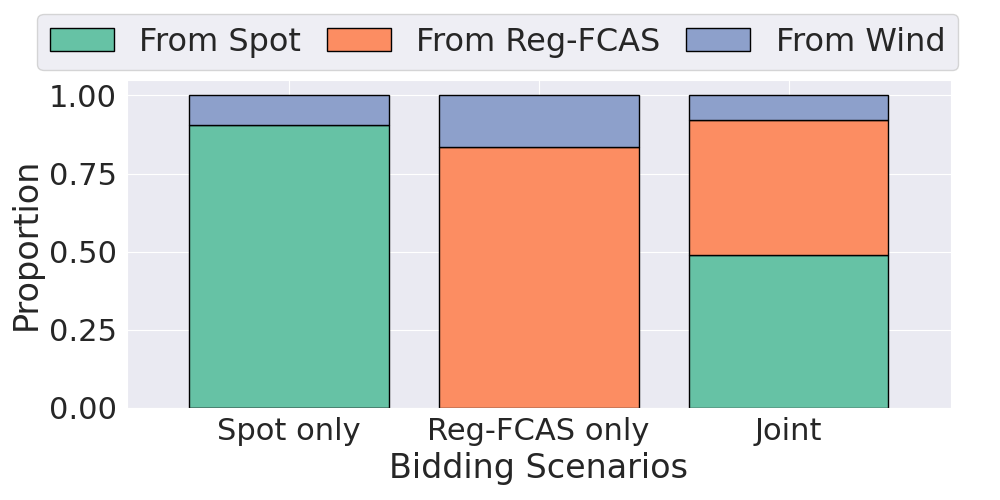}
    \caption{Composition of the BESS energy source.}
    \label{fig:BESS_power_source}
\end{figure}

The ability of the wind-BESS coupled system to reduce onsite wind curtailment is closely linked to its revenue generation. This is because reducing wind curtailment can help the wind farm, to some extent, avoid financial penalties for failing to meet dispatch targets and potentially become a source of power for charging the battery, which can be exported to the power grid in the BESS's bids. In Table \ref{tab:wc_responses_benchmark_comp}, we compare the wind curtailment response capability of the P\&O benchmark and our AggJointDRL. The results show that the AggJointDRL performs significantly better in utilizing curtailed wind energy. We also observe that the curtailed wind is a considerable source of power for charging the BESS, as illustrated in Fig. \ref{fig:BESS_power_source}. Specifically, the curtailed wind accounts for approximately 10\% of the total charged energy of the BESS in spot market trading, whereas, 16\% in Regulation FCAS, and 8\% in joint markets.

\begin{table*}[!t]
    \centering
    % \vspace*{-6mm}
    \caption{The aggregated revenue (in AU\$) with (i.e. coupled)/without (i.e., uncoupled) wind curtailment management in all three bidding scenarios.}
    \begin{tabular}{|c|c|c|c|c|}
    \cline{3-5}
    \multicolumn{2}{c|}{} & Spot & Reg. FCAS  & Joint \\
    \hline
    \multirow{3}{*}{Wind} & Uncoupled & $945,980$ & $530,150$ & $1,068,566$\\
    \cline{2-5}
    & \textbf{Coupled} & $979,784$ & $540,690$ & $1,098,686$\\
    \cline{2-5}
    & Boost & $\bm{4\%}$ & $\bm{2\%}$ & $\bm{3\%}$\\
    \hline
    \multirow{3}{*}{BESS} & Uncoupled & $94,126$ & $252,815$ & $530,219$\\
    \cline{2-5}
    & \textbf{Coupled} & $113,292$ & $247,759$ & $524,917$\\
    \cline{2-5}
    & Boost & $\bm{20\%}$ & $-2\%$ & $-1\%$\\
    \hline
    \multirow{3}{*}{Total} & Uncoupled & $1,040,106$ & $782,965$ & $1,598,785$\\
    \cline{2-5}
    & \textbf{Coupled} & $1,093,076$ & $788,449$ & $1,623,603$\\
    \cline{2-5}
    & Boost & $\bm{5\%}$ & $\bm{1\%}$ & $\bm{2\%}$\\
    \hline
    \end{tabular}
    \label{tab:evaluation_results_wwo_wc}
    % \vspace*{-2mm}
\end{table*}

To further assess the economic benefits of wind curtailment management, we have trained and evaluated the performance of a co-located wind farm and BESS as two independent players (i.e., uncoupled without wind curtailment mitigation) and compared the results to those obtained using our AggJointDRL strategy. The results are presented in Table \ref{tab:evaluation_results_wwo_wc}. In the spot market, the introduction of wind curtailment reduction leads to improved bidding performance for both the wind farm and the BESS. However, in the Regulation FCAS and the joint markets, while the wind farm still benefits from the additional financial returns brought by wind curtailment reduction, the revenue generated by the BESS decreases by 2\% and 1\%, respectively. This suggests that the BESS may face trade-offs between maximizing its own revenue and contributing to the overall financial success of the wind-battery system through wind curtailment management.

The increased revenue for the BESS in the spot market is due to the fact that using onsite curtailed wind energy is free, while charging at non-negative spot prices results in revenue losses. However, with the presence of the Regulation FCAS market, providing RL-FCAS services becomes a more favorable option than drawing curtailed wind power from the onsite wind farm due to the uncertainty of wind curtailment occurrence. As a result, responding to wind curtailment slightly diminishes the profitability of the BESS in the Regulation FCAS and the joint markets, as shown in Table \ref{tab:evaluation_results_wwo_wc}. Despite these minor revenue declines for the BESS, wind curtailment management improves the overall revenue of the wind-battery system in all three bidding scenarios, suggesting that the flexible and dynamic coordination between generation and storage is crucial to the profitability of a co-located renewable-battery coupled system. A case study in Section \ref{subsubsec:exp_wc_reduction_case_stduy_Spot_market} further examines how the BESS balances the trade-off between wind curtailment reduction and energy arbitrage in the spot market.

\begin{figure}[!t]
    \centering
    \includegraphics[width=0.8\linewidth]{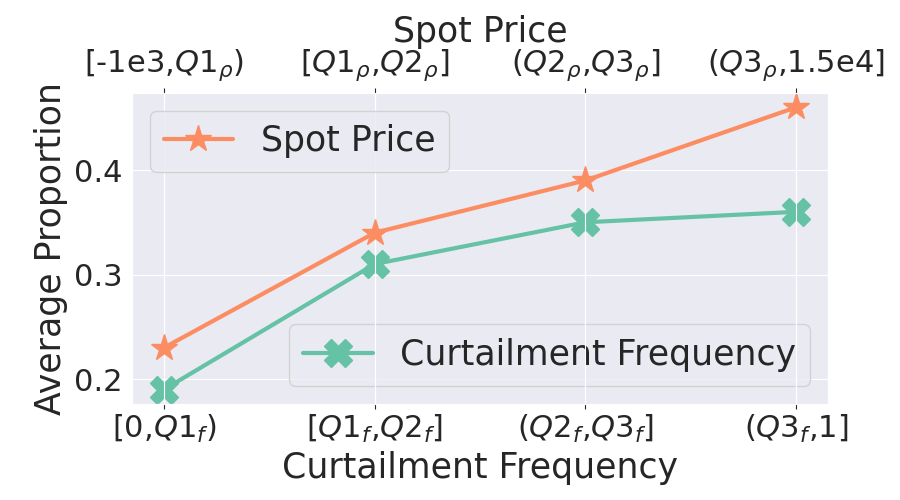}
    \caption{The impact of spot price fluctuations and wind curtailment frequencies on the bidding behavior of the BESS.}
    \label{fig:wc_case_study_Spot}
\end{figure}

\subsubsection{A Case Study in the spot market} \label{subsubsec:exp_wc_reduction_case_stduy_Spot_market}
As the charging action of the BESS can vary in response to market conditions and the availability of onsite wind curtailment, we have analyzed the relationship between spot prices and the amount of energy drawn from wind curtailment. To better understand these dynamics, we have grouped the spot prices into quartiles, $Q1_\rho$, $Q2_\rho$, and $Q3_\rho$. The results, presented in Figure \ref{fig:wc_case_study_Spot}, show that the BESS follows the principle of arbitrage and primarily sources power from the spot market when prices are low. However, when spot prices are high, the BESS tends to prioritize curtailed wind energy, as charging at these elevated prices can result in significant financial losses. While using curtailed energy as a source for charging the BESS does not incur any direct costs, it is important to note that the availability of curtailed energy is not guaranteed. In other words, the use of curtailed energy is dependent on the amount of excess energy that is generated by the wind farm and not utilized by the grid. If the amount of excess energy is insufficient, the BESS will not be able to charge using curtailed energy. Therefore, it is important for the BESS operator to consider the availability of curtailed energy in addition to the cost of sourcing power from the spot market or other sources when making charging decisions.

The decision-making process for charging the BESS is also significantly impacted by the frequency of wind curtailment, denoted as $f_t^\text{WC}$. To explore the effects of wind curtailment frequency on BESS operations, we divide the frequency into quartiles, $Q1_f$, $Q2_f$, and $Q3_f$, and examine the BESS behavior at each level. Our analysis reveals that at the lowest quartile of wind curtailment frequency, the BESS draws approximately 19\% of curtailed wind energy. As the frequency of wind curtailment increases, the BESS charges more from the onsite wind farm to reduce curtailment until reaching a plateau of approximately 36\%, as depicted in Figure \ref{fig:wc_case_study_Spot}. This leveling-off differs from that observed at high spot prices, where the BESS continues to acquire over 50\% of power from the spot market to charge the BESS. This discrepancy can be attributed to the uncertainty associated with wind curtailment, as curtailment may not necessarily occur even when the likelihood is high.

\section{Conclusion and Future Works} \label{sec:conclusions}

Our study intertwines renewable energy and artificial intelligence (AI), and presents a novel deep reinforcement learning-based bidding strategy for co-located wind-battery systems. This approach tackles key challenges in wind curtailment management and market bidding, while highlighting the potential of AI techniques, in particular, deep reinforcement learning, in driving sustainable energy solutions.

In summary, our study shows effective coordination between a co-located wind farm and BESS can significantly improve their joint bidding performance in the spot and regulation FCAS markets. To fully realize the potential of this wind-BESS system, we propose a model-free, DRL-based real-time bidding strategy called AggJointDRL. This strategy aims to optimize financial rewards by balancing BESS market bidding with wind curtailment reduction. Our analysis demonstrates that participating in the joint market significantly enhances the viability of the wind-battery system, with bidding outcomes surpassing those achieved through individual market participation. Additionally, our proposed strategy outperforms the P\&O benchmark in terms of both execution time and financial performance for both the wind farm and the BESS. We also show that onsite, otherwise curtailed wind power can be a valuable source for charging the BESS. Effective wind curtailment management leads to additional financial returns for the wind-battery system, highlighting the importance of flexible and dynamic coordination between generators and storage assets in achieving profitability. In examining the operational dynamics of the BESS under various wind curtailment frequencies and market conditions, we find that the BESS tends to use more curtailed wind energy as the spot price and wind curtailment frequency increase. These findings contribute to real-world applications by helping wind farm owners optimize their market participation strategies and guiding decisions on deploying battery energy storage systems. The successful implementation of our proposed strategy could also encourage co-location of generation and storage assets, aligning with government policies to unlock broader system benefits in energy system planning.

In the future, we plan to expand upon our joint bidding strategy for the wind-battery system by taking transmission constraints into consideration. However, a central challenge in this endeavor involves transitioning to a highly complex price-maker model, as the co-located wind-battery system may face a more competitive market environment. Additionally, we will continue to work on improving the accuracy of wind farm availability updates as a way to manage wind curtailment more efficiently.

\section*{Acknowledgments}
This work has been supported in part by the FIT Academic Funding of Monash University and the Australian Research Council (ARC) Discovery Early Career Researcher Award (DECRA) under Grant DE230100046.

\section*{Declaration of competing interest}
The authors declare that they have no known competing financial interests or personal relationships that could have appeared to influence the work reported in this paper. 

\section*{Data availability}
Data will be made available on request.

\bibliographystyle{elsarticle-num} 
\bibliography{ref}

\begin{thebibliography}{10}
\expandafter\ifx\csname url\endcsname\relax
  \def\url#1{\texttt{#1}}\fi
\expandafter\ifx\csname urlprefix\endcsname\relax\def\urlprefix{URL }\fi
\expandafter\ifx\csname href\endcsname\relax
  \def\href#1#2{#2} \def\path#1{#1}\fi

\bibitem{IPCC2022}
IPCC, {Climate Change 2022: Mitigation of Climate Change. Contribution of
  Working Group III to the Sixth Assessment Report of the Intergovernmental
  Panel on Climate Change}, Cambridge University Press, Cambridge, UK and New
  York, NY, USA, 2022.
\newblock \href {https://doi.org/10.1017/9781009157926}
  {\path{doi:10.1017/9781009157926}}.

\bibitem{CEC2022}
CEC, Clean Energy Australia Report, Clean Energy Council, 2022.

\bibitem{aemo_future_gen}
AEMO, {Australian National Electricity Market Generation Information},
  Australian Energy Market Operator, 2022.

\bibitem{burke2011}
D.~J. Burke, M.~J. O'Malley, Factors influencing wind energy curtailment, IEEE
  Transactions on Sustainable Energy 2~(2) (2011) 185--193.
\newblock \href {https://doi.org/10.1109/TSTE.2011.2104981}
  {\path{doi:10.1109/TSTE.2011.2104981}}.

\bibitem{AEMOSysPlan2022}
AEMO, Integrated System Plan for the National Electricity Market., {Australian
  Energy Market Operator}, 2022.

\bibitem{sun2017}
Y.~Sun, J.~Zhong, Z.~Li, W.~Tian, M.~Shahidehpour, Stochastic scheduling of
  battery-based energy storage transportation system with the penetration of
  wind power, in: 2017 IEEE Power \& Energy Society General Meeting, 2017, pp.
  1--1.
\newblock \href {https://doi.org/10.1109/PESGM.2017.8273785}
  {\path{doi:10.1109/PESGM.2017.8273785}}.

\bibitem{dui2018}
X.~Dui, G.~Zhu, L.~Yao, Two-stage optimization of battery energy storage
  capacity to decrease wind power curtailment in grid-connected wind farms,
  IEEE Transactions on Power Systems 33~(3) (2018) 3296--3305.
\newblock \href {https://doi.org/10.1109/TPWRS.2017.2779134}
  {\path{doi:10.1109/TPWRS.2017.2779134}}.

\bibitem{zhang2018}
Z.~Zhang, Y.~Zhang, Q.~Huang, W.-J. Lee, Market-oriented optimal dispatching
  strategy for a wind farm with a multiple stage hybrid energy storage system,
  CSEE Journal of Power and Energy Systems 4~(4) (2018) 417--424.
\newblock \href {https://doi.org/10.17775/CSEEJPES.2018.00130}
  {\path{doi:10.17775/CSEEJPES.2018.00130}}.

\bibitem{akbari2019}
E.~Akbari, R.-A. Hooshmand, M.~Gholipour, M.~Parastegari, Stochastic
  programming-based optimal bidding of compressed air energy storage with wind
  and thermal generation units in energy and reserve markets, Energy 171 (2019)
  535--546.
\newblock \href {https://doi.org/https://doi.org/10.1016/j.energy.2019.01.014}
  {\path{doi:https://doi.org/10.1016/j.energy.2019.01.014}}.

\bibitem{xie2021}
Y.~Xie, W.~Guo, Q.~Wu, K.~Wang, Robust mpc-based bidding strategy for wind
  storage systems in real-time energy and regulation markets, International
  Journal of Electrical Power \& Energy Systems 124 (2021) 106361.
\newblock \href {https://doi.org/https://doi.org/10.1016/j.ijepes.2020.106361}
  {\path{doi:https://doi.org/10.1016/j.ijepes.2020.106361}}.

\bibitem{weron2014}
R.~Weron, Electricity price forecasting: A review of the state-of-the-art with
  a look into the future, International Journal of Forecasting 30~(4) (2014)
  1030--1081.
\newblock \href
  {https://doi.org/https://doi.org/10.1016/j.ijforecast.2014.08.008}
  {\path{doi:https://doi.org/10.1016/j.ijforecast.2014.08.008}}.

\bibitem{alanazi2017}
A.~Alanazi, A.~Khodaei, Optimal battery energy storage sizing for reducing wind
  generation curtailment, in: 2017 IEEE Power \& Energy Society General
  Meeting, 2017, pp. 1--5.
\newblock \href {https://doi.org/10.1109/PESGM.2017.8274599}
  {\path{doi:10.1109/PESGM.2017.8274599}}.

\bibitem{saber2019}
H.~Saber, M.~Moeini-Aghtaie, M.~Ehsan, M.~Fotuhi-Firuzabad, A scenario-based
  planning framework for energy storage systems with the main goal of
  mitigating wind curtailment issue, International Journal of Electrical Power
  \& Energy Systems 104 (2019) 414--422.
\newblock \href {https://doi.org/10.1016/j.ijepes.2018.07.020}
  {\path{doi:10.1016/j.ijepes.2018.07.020}}.

\bibitem{nikoobahkt2020}
A.~Nikoobakht, J.~Aghaei, M.~Shafie-Khah, J.~P.~S. Catalão, Minimizing wind
  power curtailment using a continuous-time risk-based model of generating
  units and bulk energy storage, IEEE Transactions on Smart Grid 11~(6) (2020)
  4833--4846.
\newblock \href {https://doi.org/10.1109/TSG.2020.3004488}
  {\path{doi:10.1109/TSG.2020.3004488}}.

\bibitem{gill2012}
S.~Gill, G.~W. Ault, I.~Kockar, The optimal operation of energy storage in a
  wind power curtailment scheme, in: 2012 IEEE Power and Energy Society General
  Meeting, 2012, pp. 1--8.
\newblock \href {https://doi.org/10.1109/PESGM.2012.6344705}
  {\path{doi:10.1109/PESGM.2012.6344705}}.

\bibitem{khaloie2020}
H.~Khaloie, A.~Abdollahi, M.~Shafie-khah, A.~Anvari-Moghaddam, S.~Nojavan,
  P.~Siano, J.~P. Catalão, Coordinated wind-thermal-energy storage offering
  strategy in energy and spinning reserve markets using a multi-stage model,
  Applied Energy 259 (2020) 114168.
\newblock \href
  {https://doi.org/https://doi.org/10.1016/j.apenergy.2019.114168}
  {\path{doi:https://doi.org/10.1016/j.apenergy.2019.114168}}.

\bibitem{urgaokar2011}
R.~Urgaonkar, B.~Urgaonkar, M.~J. Neely, A.~Sivasubramaniam,
  \href{https://doi.org/10.1145/1993744.1993766}{Optimal power cost management
  using stored energy in data centers}, in: Proceedings of the ACM SIGMETRICS
  Joint International Conference on Measurement and Modeling of Computer
  Systems, SIGMETRICS '11, Association for Computing Machinery, New York, NY,
  USA, 2011, p. 221–232.
\newblock \href {https://doi.org/10.1145/1993744.1993766}
  {\path{doi:10.1145/1993744.1993766}}.
\newline\urlprefix\url{https://doi.org/10.1145/1993744.1993766}

\bibitem{qin2016}
J.~Qin, Y.~Chow, J.~Yang, R.~Rajagopal, Distributed online modified greedy
  algorithm for networked storage operation under uncertainty, IEEE
  Transactions on Smart Grid 7~(2) (2016) 1106--1118.
\newblock \href {https://doi.org/10.1109/TSG.2015.2422780}
  {\path{doi:10.1109/TSG.2015.2422780}}.

\bibitem{mo2021}
Y.~Mo, Q.~Lin, M.~Chen, S.-Z.~J. Qin,
  \href{https://doi.org/10.1145/3447555.3464857}{Optimal online algorithms for
  peak-demand reduction maximization with energy storage}, in: Proceedings of
  the Twelfth ACM International Conference on Future Energy Systems, e-Energy
  '21, Association for Computing Machinery, New York, NY, USA, 2021, p.
  73–83.
\newblock \href {https://doi.org/10.1145/3447555.3464857}
  {\path{doi:10.1145/3447555.3464857}}.
\newline\urlprefix\url{https://doi.org/10.1145/3447555.3464857}

\bibitem{yang2020}
J.~Yang, M.~Yang, M.~Wang, P.~Du, Y.~Yu, A deep reinforcement learning method
  for managing wind farm uncertainties through energy storage system control
  and external reserve purchasing, International Journal of Electrical Power \&
  Energy Systems 119 (2020) 105928.
\newblock \href {https://doi.org/https://doi.org/10.1016/j.ijepes.2020.105928}
  {\path{doi:https://doi.org/10.1016/j.ijepes.2020.105928}}.

\bibitem{oh2020}
E.~Oh, H.~Wang, Reinforcement-learning-based energy storage system operation
  strategies to manage wind power forecast uncertainty, IEEE Access 8 (2020)
  20965--20976.
\newblock \href {https://doi.org/10.1109/ACCESS.2020.2968841}
  {\path{doi:10.1109/ACCESS.2020.2968841}}.

\bibitem{aemo2021}
AEMO, Guide to Market Clearing, Australian Energy Market Operator, 2021.

\bibitem{aemo2020}
AEMO, How the National Electricity Market works, Australian Energy Market
  Operator, 2020.

\bibitem{aemo2017}
AEMO, Five-minute Settlement: High Level Design, Australian Energy Market
  Operator, 2017.

\bibitem{aemo2015}
AEMO, Guide to Ancillary Services in the National Electricity Market,
  Australian Energy Market Operator, 2015.

\bibitem{aemo2022}
AEMO, NEM Operational Forecasting and Dispatch Handbook for wind and solar
  generators, Australian Energy Market Operator, 2022.

\bibitem{bordin2017}
C.~Bordin, H.~O. Anuta, A.~Crossland, I.~L. Gutierrez, C.~J. Dent, D.~Vigo, A
  linear programming approach for battery degradation analysis and optimization
  in offgrid power systems with solar energy integration, Renewable Energy 101
  (2017) 417--430.
\newblock \href {https://doi.org/https://doi.org/10.1016/j.renene.2016.08.066}
  {\path{doi:https://doi.org/10.1016/j.renene.2016.08.066}}.

\bibitem{anwar2022}
M.~Anwar, C.~Wang, F.~de~Nijs, H.~Wang, Proximal policy optimization based
  reinforcement learning for joint bidding in energy and frequency regulation
  markets, IEEE Power \& Energy Society General Meeting (PESGM) (2022).

\bibitem{fujimoto2018}
S.~Fujimoto, H.~Hoof, D.~Meger, Addressing function approximation error in
  actor-critic methods, in: International Conference on Machine Learning, 2018,
  pp. 1582--1591.

\bibitem{goodfellow2016}
I.~Goodfellow, Y.~Bengio, A.~Courville, Deep Learning, MIT Press, 2016.

\bibitem{sutton2018}
R.~S. Sutton, A.~G. Barto, Reinforcement Learning: An Introduction, 2nd
  Edition, The MIT Press, 2018.

\bibitem{hasselt2010}
H.~Hasselt, Double q-learning, in: J.~Lafferty, C.~Williams, J.~Shawe-Taylor,
  R.~Zemel, A.~Culotta (Eds.), Advances in Neural Information Processing
  Systems, Vol.~23, Curran Associates, Inc., 2010, pp. 1--9.

\bibitem{aemo_data}
AEMO, AEMO Nemweb data, Australian Energy Market Operator, 2022.

\bibitem{mitchell2011}
S.~Mitchell, M.~OSullivan, I.~Dunning, Pulp: a linear programming toolkit for
  python, The University of Auckland, Auckland, New Zealand 65 (2011).

\bibitem{schulman2017}
J.~Schulman, F.~Wolski, P.~Dhariwal, A.~Radford, O.~Klimov, Proximal policy
  optimization algorithms, CoRR abs/1707.06347 (2017).

\bibitem{lillicrap2016}
T.~P. Lillicrap, J.~J. Hunt, A.~Pritzel, N.~Heess, T.~Erez, Y.~Tassa,
  D.~Silver, D.~Wierstra, Continuous control with deep reinforcement learning
  (2015).
\newblock \href {https://doi.org/10.48550/ARXIV.1509.02971}
  {\path{doi:10.48550/ARXIV.1509.02971}}.

\bibitem{haarnoja2018}
T.~Haarnoja, A.~Zhou, P.~Abbeel, S.~Levine, Soft actor-critic: Off-policy
  maximum entropy deep reinforcement learning with a stochastic actor, in:
  Proceedings of the 35th International Conference on Machine Learning, Vol.~80
  of Proceedings of Machine Learning Research, PMLR, 2018, pp. 1861--1870.

\end{thebibliography}
\end{document}